\documentclass[notitlepage,prd,twocolumn,showpacs,preprintnumbers,nofootinbib,superscriptaddress,floatfix,tightenlines]{revtex4-2}

\usepackage{graphicx}
\usepackage{dcolumn}
\usepackage{bm}
\usepackage{hyperref}
\usepackage{amsmath, amssymb, amsfonts, amsbsy, mathrsfs, bbm}
\usepackage{booktabs}
\usepackage{xargs}
\usepackage[colorinlistoftodos,prependcaption,textsize=footnotesize]{todonotes}
\newcommandx{\DM}[2][1=]{\todo[linecolor=orange,backgroundcolor=orange!25,bordercolor=orange,#1]{DM: #2}}
\newcommandx{\KB}[2][1=]{\todo[linecolor=purple,backgroundcolor=purple!25,bordercolor=purple,#1]{KB: #2}}
\newcommandx{\PH}[2][1=]{\todo[linecolor=yellow,backgroundcolor=yellow!25,bordercolor=yellow,#1]{PH: #2}}
\newcommandx{\TODO}[2][1=]{\todo[linecolor=red,backgroundcolor=red!25,bordercolor=red,caption={TODO},#1]{#2}}

\usepackage{comment}
\begin{document}

\title{Complex Langevin kernels for real-time SU(N) \\ lattice gauge theory in 3+1 dimensions}

\author{Kirill Boguslavski} 
\affiliation{SUBATECH UMR 6457 (IMT Atlantique, Université de Nantes, IN2P3/CNRS), 4 rue Alfred Kastler, 44307 Nantes, France}

\author{Paul Hotzy} 
\author{David I.~M\"uller} 
\affiliation{Institute for Theoretical Physics, TU Wien, Wiedner Hauptstraße 8-10, 1040 Vienna, Austria}

\begin{abstract}
We investigate complex Langevin (CL) dynamics for real-time SU($N$) gauge theories in 3+1 dimensions and introduce new kernel modifications tailored for lattice gauge theory (LGT). Using the anisotropic kernel as a baseline, we demonstrate that on a rectangular regularized Schwinger-Keldysh time contour, stable simulations are possible for pure SU(2) gauge theory up to a real-time extent of $2\beta$ at a bare coupling $g=0.5$, where $\beta$ denotes the inverse temperature. We show that a contour-dependent phase kernel extends this reach to $3\beta$, while a local, field-dependent kernel based on traced plaquettes achieves $3.5\beta$. These constructions represent the first application of such kernels to LGTs, especially in 3+1 dimensions. To validate the approach, we generalize the CL correctness criterion based on the drift magnitude to kernelled dynamics, providing a practical tool to assess correct convergence. Our results substantially advance the applicability of CL to real-time simulations in Minkowski space, establishing a framework for controlled studies of non-Abelian gauge theories at unprecedented time scales. A forthcoming paper applies these developments to the computation of unequal-time correlation functions.
\end{abstract}
\date{\today}

\maketitle

\section{Introduction}

Real-time dynamics in quantum field theory describe a wide range of physical phenomena, from the extreme conditions in the early universe to modern collider experiments. Governed by quantum chromodynamics (QCD), the time evolution of the quark-gluon plasma (QGP) formed at RHIC and the LHC is particularly difficult to study, as it requires access to transport properties and spectral information that are inherently real-time and non-perturbative \cite{Busza:2018rrf,Apolinario:2022vzg}. While weak-coupling perturbative methods provide insight at asymptotically high energies, reliable calculations at experimentally relevant scales demand fully non-perturbative approaches.

Lattice QCD offers the most robust non-perturbative framework for studying strongly interacting gauge theories \cite{Gattringer:2010zz}. However, standard lattice simulations rely on a Euclidean formulation, which prevents direct access to real-time observables. Reconstructing Minkowski-space information from Euclidean correlators is an ill-posed inverse problem \cite{Asakawa:2000tr,Meyer:2011gj,Rothkopf:2022ctl}. Attempting instead to formulate lattice dynamics directly in Minkowski time leads to a highly oscillatory path-integral weight \cite{Aarts:2026uiu,Kanwar:2021tkd}. This becomes evident in the Schwinger-Keldysh (SK) contour approach \cite{Schwinger:1960qe,Keldysh:1964ud} to real-time quantum field theory, where the path integral acquires complex weights that lead to cancellations between contributions with rapidly varying phases. This real-time sign (or phase) problem renders standard importance sampling ineffective \cite{Gattringer:2016kco,Kanwar:2021tkd}.

The complex Langevin (CL) method \cite{Parisi:1983complex,Damgaard:1987rr,Aarts:2026uiu} offers a promising approach. Extending the stochastic quantization of quantum field theory \cite{Parisi:1980ys} to complex actions opens the possibility of performing simulations directly on complex-time contours and, in particular, on the SK time contour. Despite notable successes in real-time scalar systems, QCD at finite chemical potential, and other lower-dimensional theories with complex actions \cite{Berges:2005yt, Berges:2006xc, Alvestad:2023jgl, Mandl:2026ngb, Joseph:2020gdh, Boguslavski:2024yto}, its application to non-Abelian SU($N$) lattice gauge theories (LGT) on complex time paths \cite{Berges:2006xc, Berges:2007nr, Aarts:2017hqp} has been obstructed by runaway trajectories and, most importantly, convergence to incorrect results when the underlying dynamics become insufficiently controlled \cite{Aarts:2009uq}. These problems generally emerge in the CL approach, such as for the study of QCD at finite chemical potential. 
To address these issues, various stabilization mechanisms, such as gauge cooling \cite{Seiler:2012wz}, dynamical stabilization \cite{Attanasio:2018rtq}, or kernel modifications \cite{Okamoto:1988ru,Okano:1991tz}, have been proposed. An important conceptual advancement has been achieved by the formulation of correctness criteria \cite{Aarts:2009uq, Nagata:2016vkn, Seiler:2023kes, Hansen:2024kjm, Mandl:2025mav, Mandl:2026ebc} that can (in)validate the correct convergence and robustness of the obtained CL simulation results, rendering the CL method more practical.
Real-time simulations of LGTs finally became possible with a new (anisotropic) kernel that exploits the kernel freedom of CL and allows for simulations even in higher dimensions \cite{Boguslavski:2022dee,Boguslavski:2023unu}.

Related progress has been made in real-time scalar and lower-dimensional systems, where stable implicit solvers, kernel modifications, and machine-learning-assisted kernel optimization have enabled simulations of interacting scalar field theory in up to $1+1$ dimensions \cite{Alvestad:2021hsi, Alvestad:2022abf, Lampl:2023xpb, Alvestad:2023jgl}. 
Complementary non-perturbative approaches to real-time scalar fields include contour deformation \cite{Alexandru:2016gsd, Alexandru:2017lqr} and real-time formulations of the functional renormalization group \cite{Pawlowski:2015mia,Huelsmann:2020xcy}, whereas related generative approaches based on physics-informed kernels have recently been proposed for lattice field theory \cite{Ihssen:2025ybn} with a first real-time demonstration in quantum mechanics \cite{Ihssen:2026njd}. These developments point to a broader role of kernel-based transformations and learned structures in approaching real-time and complex-weight problems.

To tackle the problem of real-time non-Abelian LGTs with the CL method, we introduced the anisotropic kernel tailored to gauge theories and the geometry of complex-time contours \cite{Boguslavski:2022dee}. 
This enabled the first stable extraction of unequal-time correlation functions in $3+1$-dimensional Yang-Mills theory for real-time extents up to $t_{\max}=1.5\beta$ for the simulation parameters considered \cite{Boguslavski:2023unu}, where $\beta=1/T$ is the inverse temperature. While this constituted a significant advancement, the method becomes unstable for longer real-time extents or larger gauge couplings unless the lattice spacing anisotropy is drastically increased, which quickly becomes costly. Extending simulations to larger $t_{\max}$ is essential for accessing physically relevant observables, such as transport coefficients and spectral functions at low frequencies \cite{Meyer:2011gj, Altenkort:2022yhb}.

The aim of the present work is to develop new kernel modifications, both field-independent and field-dependent, that further enhance the stability of the CL process for real-time LGT. These constructions exploit the specific geometric information on the studied systems and the complex-time contours as well as local gauge-invariant information in order to achieve correctly converged results over a wider parameter space, in particular at significantly larger real time extents $t_{\max}$.

With these new kernels and a rectangular SK contour, we achieve a substantial extension of the accessible real-time window as compared to our previous work \cite{Boguslavski:2023unu}. The rectangular SK contour alone extends to $t_{\max}=2\beta$. A contour-dependent phase kernel allows simulations up to $t_{\max}=3\beta$, while a local, field-dependent kernel based on traced plaquettes pushes this boundary further to $t_{\max}=3.5\beta$. 
This more than doubles the real-time extent achieved in Ref.~\cite{Boguslavski:2023unu}, which potentially allows for a more robust frequency-domain analysis of correlation functions of interest \cite{companion:2026}.

In addition, we generalize the criterion of correct convergence proposed by Nagata et al.~\cite{Nagata:2016vkn} for complex Langevin dynamics to encompass kernelled evolutions, thereby providing a practical \emph{a posteriori} validation tool for our simulations. This extension is particularly relevant in view of recent results demonstrating that kernels can alter the integration cycles sampled by CL dynamics and that conventional convergence diagnostics alone need not be sufficient to identify the desired result \cite{Hansen:2024kjm,Mandl:2025mav}.

This paper is organized as follows. We discuss the complex Langevin method applied to real-time LGT and CL kernels in Sec.~\ref{sec:CL_method}. In Sec.~\ref{sec:coc_gamma}, we extend the criterion of correctness by Nagata et al.~to kernelled complex Langevin simulations. We then introduce and verify field-independent phase kernels in Sec.~\ref{sec:field-indep_kernels} and field-dependent kernels in Sec.~\ref{sec:field-dep_kernels}, before finally concluding in Sec.~\ref{sec:conclusion}.

\section{Complex Langevin method}
\label{sec:CL_method}

    \begin{figure}[t]
    \centering
    \includegraphics[width=0.5\textwidth]{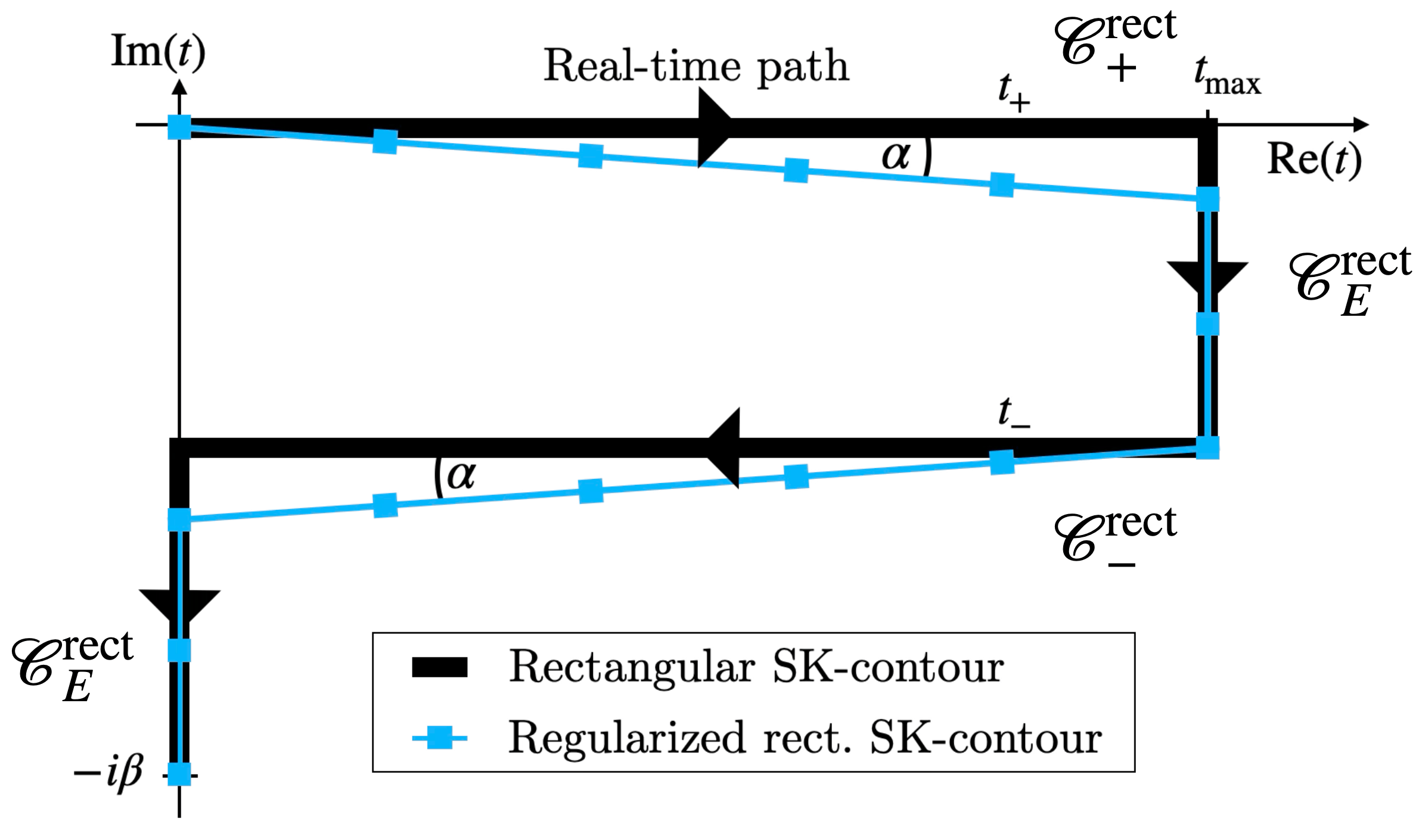}
    \caption{
        Schematic illustration of the (rectangular) Schwinger–Keldysh (SK) contour in the complex time plane. The black curve represents the undeformed contour, while the blue curve shows its regularized, lattice-discretized version with tilt angle $\alpha$, complex time step $a_0$, and two (vertical) Euclidean branches of approximate length $\beta/2$. The sketch is not to scale; the number and spacing of points do not correspond to specific simulation parameters.
        \label{fig:sk_contour}
    }
    \end{figure}

    We apply the complex Langevin (CL) method to non-Abelian gauge fields defined along a (rectangular) complex Schwinger–Keldysh (SK) time contour $\mathscr{C} $, as illustrated in Fig.~\ref{fig:sk_contour}. This enables us to directly access real-time properties of gauge fields.
    The dynamics are governed by the Yang–Mills action evaluated on this contour
    \begin{align} \label{eq:ym_action}
        S_\mathrm{SK} = - \frac{1}{4} \int_{\mathscr C} d^{d+1}x\, F^{\mu \nu}_a F_{\mu \nu}^a,
    \end{align}
    where the field strength tensor is defined as
    \begin{align}
        F_{\mu \nu}^a = \partial_\mu A^a_\nu - \partial_\nu A^a_\mu - g f^{abc} A^b_\mu A^c_\nu.
    \end{align}
    Here, $f^{abc}$ are the antisymmetric structure constants of the Lie algebra $\mathfrak{su}(N) $, and $g$ denotes the gauge coupling. We use the Einstein summation convention for the (Minkowski) spacetime indices $\mu, \nu = 0, 1, \dots, d$ and color indices $a, b = 1, \dots, N^2 - 1 $, unless the quantity at question has an explicit index.
    
    Our objective is to compute real-time expectation values of gauge-invariant observables. 
    In the SK formalism, the path integral takes the form
    \begin{align} \label{eq:pi_gauge}
        &\langle \mathcal{O} \rangle = \frac{1}{Z} \int \mathcal{D}A \, e^{iS_\mathrm{SK}[A]}\, \mathcal{O} \nonumber \\
        &= \frac{1}{Z}\! \int \! \! \mathcal{D}A_E e^{-S_E} \!\int \!\!\mathcal{D}A_+ \mathcal{D}A_- e^{i S^+_M-iS^-_M}\, \mathcal{O},
    \end{align}
    where $S^\pm_M = S_\mathrm{SK}|_{\mathscr{C}_\pm}$ denotes the Minkowski action along the real-time paths $\mathscr{C}_\pm$. 
    Thermal equilibrium is imposed by integrating along the Euclidean paths $\mathscr{C}_E$ with Euclidean action $S_E = -iS_\mathrm{SK}|_{\mathscr{C}_E}$ and periodic boundary conditions $A^a_\mu(t = 0) = A^a_\mu(t = -i\beta) $, with the inverse temperature $\beta = 1/T$.

    Instead of evaluating the path integral \eqref{eq:pi_gauge} directly, we employ stochastic quantization \cite{Parisi:1983complex,Damgaard:1987rr,Aarts:2026uiu, Parisi:1980ys} and solve the corresponding Langevin equation numerically. To apply this method to $\mathrm{SU}(N)$ gauge theories along complex time contours, one must complexify the gauge fields. The algebra is extended as $\mathfrak{su}(N) \to \mathfrak{sl}(N, \mathbb{C}) $. The resulting {\em complex} Langevin equation for the gauge field components then reads
    \begin{align} \label{eq:cle}
        \frac{\partial A^a_\mu}{\partial \theta}(\theta, t, x) = i \left. \frac{\delta S_\mathrm{SK}}{\delta A^a_\mu(t, x)} \right|_{A = A(\theta)} + \eta^a_\mu(\theta, t, x),
    \end{align}
    where $\theta$ denotes the Langevin time, and the Gaussian noise is independent for each color and Lorentz index and satisfies
    \begin{align}
        \langle \eta^a_\mu(\theta, t, x) \rangle &= 0, \\
        \!\!\langle \eta^a_\mu(\theta, t, x)\, \eta^b_\nu(\theta', t', x') \rangle &= 2 \delta(\theta - \theta') \times \nonumber \\
        &\quad\delta(t - t')\delta^{(d)}(x - x') \delta^{ab} \delta_{\mu \nu}.
    \end{align}
    The complex structure of the SK contour induces subtleties in the definition of the noise correlator due to its complex argument structure. These issues are addressed by a suitable parametrization of the contour, as discussed in Ref.~\cite{Boguslavski:2022dee}, which motivates an anisotropic kernel. 
    
    The CL equation provides correct results if the field configurations at large Langevin time $\theta > \theta_0$ are distributed according to a real probability density $P$ that reproduces the original complex weight in Eq.~\eqref{eq:pi_gauge}. Under this condition, expectation values are computed as Langevin time averages
    \begin{align}
        \langle \mathcal{O} \rangle = \lim_{\Theta \to \infty} \frac{1}{\Theta} \int_{\theta_0}^{\theta_0 + \Theta} d\theta\, \mathcal{O}(U(\theta)).
    \end{align}

    \subsection{CL method for lattice gauge theories}
    \label{sec:CL_lattice_gauge}

    To exactly preserve gauge-covariance of the theory, we solve the CL equation \eqref{eq:cle} on a hyper-cubic space-time lattice for the link fields 
    \begin{align}
        U_{x,\mu} \simeq \exp{\left[i ga_\mu t^a  A^a_{\mu}(x+\hat \mu/2\big)\right]},
    \end{align}
    and discretize the Yang-Mills action using the Wilson plaquette action
    \begin{align}\label{eq:wilson_action}
        S_\mathrm{w} = \frac{1}{g^2} \sum_{x}\sum_{\mu, \nu \neq \mu} \rho_{x,\mu\nu} \mathrm{Tr} \left[U_{x,\mu\nu} - \mathbbm{1}\right],
    \end{align}
    where the plaquette variables 
    \begin{align}
        U_{x,\mu\nu} = U_{x,\mu}\, U_{x+\hat{\mu}, \nu}\, U^{-1}_{x+\hat{\nu},\mu}\, U^{-1}_{x,\nu}
    \end{align}
    form the smallest Wilson loops in terms of link fields. The sum runs over all lattice sites $x$ and both directions $\mu, \nu$ leaving out terms with $\mu = \nu$.
    The prefactors $\rho_{x,0i} = -a_s / a_0(t)$ and $\rho_{x,ij} = a_0(t)/a_s$ entering the Wilson action depend on the temporal and spatial lattice spacings $a_0(t)$ and $a_s$. 
    A more detailed discussion of the discretization can be found in Ref.~\cite{Boguslavski:2022dee}. 
    We note that here, we have deliberately written the dependence on the discretized lattice space-time $x$ as an index with the only exception being $a_0(t)$ to explicitly distinguish the temporal component $0$ from the dependence on the complex time contour $t$.
    
    The temporal spacing encodes the geometry of the regularized, rectangular Schwinger-Keldysh contour depicted as the blue curve in Fig.~\ref{fig:sk_contour} and is therefore time-dependent. To allow a well-defined temporal continuum limit, we regularize the discretized real-time paths by tilting them towards the imaginary time axis by an angle $\alpha$. Throughout this work, we impose that along the entire SK contour, the absolute value of the temporal spacing is constant $\vert a_0(t) \vert \approx a_s/32$, with a spatial spacing $a_s=1$. In principle, to eliminate regularization artifacts in real-time observables, one should numerically perform the limit $\alpha\to0$ as introduced in Ref.~\cite{Boguslavski:2023unu}. As we wish here to compare and assess different CL kernels, we employ a small but finite angle $\tan \alpha = 1/100$ instead, while leaving the computation of observables from $\alpha \to 0$ for a forthcoming study \cite{companion:2026}.

\subsection{Discretized CL equation with field-independent kernels} 
\label{sec:update_indep_kernel}

    The discretized CL equation corresponding to \eqref{eq:cle} reads for the Euler-Maruyama update scheme
    \begin{align} \label{eq:CL_latt_gamma_indep}
        U_{x,\mu}(\theta   +   \epsilon) &= e^{i t^a \left(\epsilon K^{\Gamma,a}_{x,\mu}[U(\theta)] + \sqrt{ \epsilon}\, \gamma_{x,\mu}\, \eta^a_{x,\mu}(\theta)\right)} \,U_{x,\mu}(\theta), \nonumber\\
        K^{\Gamma,a}_{x,\mu}[U(\theta)] &= \Gamma_{x,\mu} K_{x,\mu}^a[U(\theta)], \quad \gamma_{x,\mu} = \sqrt{\Gamma_{x,\mu}}\,,
    \end{align}
    where $\epsilon$ represents the Langevin time step.
    The drift term $K$ is given as the group derivative of $i S_\mathrm{w}$ with respect to the link $U_{x,\mu}$,
    \begin{align}
        K_{x,\mu}^a[U(\theta)]= D^a_{x,\mu}\, iS_\mathrm{w} = \frac{i}{g a_\mu}\,\frac{\delta S_\mathrm{w}}{\delta A^a_{ \mu}(x)},
    \end{align}
    which coincides with the functional derivative with respect to $g a_\mu A^a_{ \mu}(x)$.
    The Gaussian noise field satisfies
    \begin{align}
        \langle \eta^a_{x,\mu}(\theta) \rangle &= 0, \\
        \langle \eta^a_{x,\mu}(\theta) \eta^b_{y,\nu}(\theta') \rangle &= 2 \delta_{\theta\theta'} \delta_{xy} \delta^{ab} \delta_{\mu\nu}.
    \end{align}
    We note that this leading-order update scheme is $\mathcal{O}(\epsilon)$.
    
    We have introduced a field-independent kernel function $\Gamma_{\mu} (x)$ in \eqref{eq:CL_latt_gamma_indep}. 
    In the real-valued (Euclidean) case,  kernels leave the stationary solution invariant. For CL, kernels generally modify the stationary solution and the convergence properties of the process. 
    In particular, the kernel freedom can be exploited to stabilize and improve the convergence properties of the CL process by rescaling different degrees of freedom independently. 
    
    In Ref.~\cite{Boguslavski:2022dee}, we have introduced and motivated the anisotropic kernel
    \begin{align} \label{eq:aniso_kernel}
        \Gamma_{x,0} = \frac{|a_0(t)|^2} {a_s^2} = \xi^{-2},
        \quad \Gamma_{x,i} = 1,
    \end{align}
    which originates from a careful contour parametrization of the complex time path. It leads to a substantial stabilizing effect by increasing the number of discretization points along the time contour, i.e., by increasing the lattice spacing anisotropy $\xi = a_s/|a_0|$. 
    Therefore, it is incorporated in all simulations conducted in this work. 
    
    Importantly, the introduction of the anisotropic kernel enabled the first extraction of unequal-time correlation functions for real-time Yang-Mills theory in physical four dimensions in Ref.~\cite{Boguslavski:2023unu}. There, a complex SK time contour with an isosceles regularization of the real-time path was used instead of the rectangular contour in Fig.~\ref{fig:sk_contour}. This allowed us to compare different correlation functions by combining fields on the forward and backward time paths at the expense of a shorter real-time extent of up to $t_{\mathrm{max}}=1.5\beta$. 
    
    In the present work, we will improve on this benchmark real-time extent by introducing several advancements. 
    The focus of our study is the development of new kernel transformations, including both field-independent and field-dependent kernels, that we test in the present work and use to further increase the real-time extent. 
    In addition, we employ the more stable rectangular SK contour of Fig.~\ref{fig:sk_contour} that already allows extending the real-time extent to $t_{\mathrm{max}}=2\beta$ without changing other parameters.

\subsection{Kernelled CL update step}
\label{sec:kernel_CL_update}

    The general kernelled form of the CL equation \eqref{eq:cle} reads
    \begin{align} \label{eq:ym_cle_kernel}
        \frac{\partial A^a_\mu}{\partial \theta}(x;\theta) =
        i& \int d^{d+1}x' \,  \Gamma^{ab}_{\mu\nu}(x,x') \left. \frac{\delta S}{\delta A^b_\nu(x')} \right|_{A=A(\theta)} + \nonumber \\
        & \int d^{d+1}x' \, \left. \frac{\delta \Gamma^{ab}_{\mu\nu}(x,x')}{\delta A^b_\nu(x')} \right|_{A=A(\theta)} + \nonumber \\
        &\int d^{d+1}x' \, \gamma^{ab}_{\mu\nu}(x,x') \eta_b^\nu(\theta, x'),
    \end{align}
    where the kernel $\Gamma^{ab}_{\mu\nu}(x, x')$ is assumed to be holomorphic and factorizable,
    \begin{align}
        \label{eq:Gamma_gamma}
        \Gamma^{ab}_{\mu\nu}(x,x') = \int d^{d+1}x'' \, \gamma^{ac}_{\mu\sigma}(x, x'') \gamma^{bc}_{\nu\sigma}(x', x'').
    \end{align}
    This stochastic equation can be associated with a complex Fokker-Planck equation that admits the same stationary complex weight function $\rho\propto e^{iS}$ as in Eq.~\eqref{eq:pi_gauge}. We note, however, that the real probability density $P(\mathrm{Re}(A_\mu^a), \mathrm{Im}(A_\nu^b))$ of the stochastic process is changed depending on the specific kernel, and hence, we can effectively manipulate the stability and convergence of the CL process. 

    On the lattice, we can reformulate the kernelled CL equation in terms of link fields. In this work, we focus on diagonal kernel forms, assuming
    \begin{align}
        \label{eq:Gamma_diag_gamma}
        \begin{split}
        \Gamma^{ab}_{xx',\mu\nu}[U] &= \delta^{ab} \delta_{\mu\nu} \delta_{xx'} \Gamma_{x,\mu}[U], \\
        \gamma_{x,\mu}[U] &= \sqrt{\Gamma_{x,\mu}[U]}.
        \end{split}
    \end{align}
    This diagonal form preserves all symmetries of the theory and is an essential simplification as it strongly reduces the number of parameter and functional dependencies to choose for the kernels. We will motivate and provide specific kernels in Secs.~\ref{sec:field-indep_kernels} and \ref{sec:field-dep_kernels}. 
    We also note that, in principle, taking a non-diagonal structure for $\Gamma$ could allow for more handles to further improve the convergence properties of the CL simulations, but would come at the cost of introducing more complexity to the problem.
    
    The resulting leading-order Euler-Maruyama update scheme takes the form
    \begin{align} \label{eq:link_update_gamma}
        \begin{split}
            U_{x,\mu}(\theta   +   \epsilon) =\, &e^{i t^a \left(\epsilon K^{\Gamma, a}_{x,\mu}[U(\theta)]+ \sqrt{\epsilon} \, \gamma_{x,\mu}\, \eta^a_{x,\mu}\right)} \, U_{x,\mu}(\theta),\\
            K^{\Gamma, a}_{x,\mu}[U(\theta)] = \,&  \Gamma_{x,\mu} K_{x,\mu}^a + D^a_{x,\mu}\Gamma_{x,\mu},
        \end{split}
    \end{align}
    which extends Eq.~\eqref{eq:CL_latt_gamma_indep} by an additional group derivative of the kernel function $D^a_{x,\mu}\Gamma_{x,\mu} = \delta \Gamma_{x,\mu}\big/\delta g a_\mu A_\mu^a(x)$.

\subsection{Higher-order solver}
\label{sec:higher_order_maintext}

For consistency with previous results and to ensure a meaningful comparison, we employ a Runge-Kutta-like stochastic solver, as described in Appendix~\ref{app:higherorder_sde}. The choice is motivated by the following considerations:
\begin{itemize}
    \item For field-dependent kernels, the Euler-Maruyama scheme converges strongly only with order $\mathcal{O}(\epsilon^{1/2})$ but still weakly with order $\mathcal{O}(\epsilon)$.
    \item For the anisotropic kernel, our earlier work used an ``improved step'' method with the higher second-order convergence $\mathcal{O}(\epsilon^2)$.
    \item However, this ``improved step'' method of our previous studies cannot be used here directly, as it does not account for the field-dependent diffusion term.
\end{itemize}
The chosen solver retains second-order convergence, $\mathcal{O}(\epsilon^2)$, in the sense of both strong and weak convergence, and is also applicable to field-dependent kernels, allowing for a direct and fair comparison across different kernel settings.

\section{Criterion of correctness for kernelled complex Langevin} \label{sec:coc_gamma}

In this section, we examine a criterion for the correctness of CL dynamics. It provides a practical, though \emph{a posteriori}, method to assess whether the probability density obtained from the CL process correctly reproduces the complex path integral weight. The criterion was introduced in Ref.~\cite{Nagata:2016vkn} and is based on the decay behavior of the probability density of the CL drift magnitude. It is equivalent to an alternative criterion of correctness based on boundary terms \cite{Scherzer:2018hid, Scherzer:2019lrh} that requires that during the Langevin evolution, there are no boundary times for (polynomial) observables, including correlation functions of the fields. 
We choose the drift magnitude criterion \cite{Nagata:2016vkn} in our situation, as it is observable-independent. 

After a short summary, we extend the correctness criterion in this section to lattice gauge theories and kernelled CL equations, with details given in Appendix \ref{app:coc}.

    \subsection{Diagnosis of correct convergence of CL}

        Following the argument of Ref.~\cite{Nagata:2016vkn}, we discuss the CL correctness criterion for a zero-dimensional dynamical variable $Z(\theta) = X(\theta) +i Y(\theta)$ described by a CL equation.
        It states that the probability density $p(u, \theta)$ of the drift magnitude $u(x + i y) := \vert K(x + i y) \vert$,
        \begin{align} 
            p(u, \theta) = \int dx \, dy \, \delta(u(x + i y) - u) \, P_{Z}(x, y; \theta), 
        \end{align}
        with respect to the real probability distribution $P_{Z}(x, y; \theta)$, has to decay at least exponentially for all Langevin times.
        This decay has to be bounded by an exponential with a finite decay rate $r_{\mathrm{conv}} > 0$, making the criterion sufficient to demonstrate the correctness of the CL evolution. 
        
        This condition is based on the comparison of the evolution of the complex weight $\rho(x, \theta)$, following a complex Fokker-Planck equation, and the real probability distribution $P_{Z}(x, y; \theta)$ of the real and imaginary parts of the dynamical variable $z$ acting on an observable during a Langevin update step $\theta \to \theta + \epsilon$. 
        It was shown in \cite{Nagata:2016vkn} that both agree if the update step for the Fokker-Planck equation of $P_{Z}(x,y; \theta) \to P_{Z}(x,y; \theta+\epsilon)$ converges uniformly with the finite convergence radius $\epsilon < r_{\mathrm{conv}}$. 
        Hence, this argument also applies to a discretized CL equation with finite Langevin time step $\epsilon$.
        This provides a practical criterion for CL simulations that can be checked a posteriori by studying the decay rate of $p(u, \theta)$.

    \subsection{Correctness criterion for lattice gauge theories} \label{sec:coc_lattice}

        In the context of real-time SU($N$) Yang-Mills theory, we define the drift magnitude as the maximal value of the drift under all (unitary) gauge transformations
        \begin{align}
        |K[{U}]|_{\infty}
        = \max_{\Omega_{x} \in SU(N)} \max_{x,\mu,a} \left\vert K_{x,\mu}^a [{U^\Omega }]\right\vert,
        \end{align}
        where ${U^\Omega}$ denotes the gauge-transformed link field configuration
        \begin{align}
        U_{x,\mu} \mapsto \Omega_{x} U_{x,\mu} \Omega_{x+\mu}^\dagger.
        \end{align}
        
        While this definition is conceptually sound, computing the global maximum over all possible gauge transformations is computationally infeasible. To overcome this, we instead estimate the drift magnitude using the following sequence of bounds
        \begin{align}
        |K[{U}]|_{\infty}
        &\leq \max_{\Omega_{x,\mu} \in SU(N)} \max_{x,\mu} \sqrt{\sum_a \left| K^a_{x,\mu}[U^\Omega] \right|^2} \nonumber \\
        &= \max_{x, \mu} \sqrt{\sum_a \left| K^a_{x,\mu}[U] \right|^2} \nonumber \\
        &= \max_{x, \mu} \sqrt{\sum_a \left( \mathrm{Re} K^a_{x,\mu}[U] \right)^2 + \left( \mathrm{Im} K^a_{x,\mu}[U] \right)^2} \nonumber \\
        &\leq \max_{x, \mu} \sum_a \left( \left| \mathrm{Re} K^a_{x,\mu}[U] \right| + \left| \mathrm{Im} K^a_{x,\mu}[U] \right| \right) \nonumber \\
        &=: N_K[{U}],
        \end{align}
        where $\mathrm{Re} K^a_{x,\mu}$ and $\mathrm{Im} K^a_{x,\mu}$ are the real and imaginary parts of the drift components. 
        In the first line, we used that the drift transforms locally as $K^{a}_{x,\mu}[{U^\Omega }] = \Omega^{ab}_x K^{b}_{x,\mu}[{U}]$, where $\Omega^{ab} = 2 \mathrm{Tr}(t^a \Omega t^b \Omega^\dagger)$ is the adjoint representation of $\Omega$. 
        Due to the equivalence of norms in finite-dimensional spaces, these bounds provide a reliable and computationally tractable measure of the drift magnitude, sufficient for analyzing the decay properties of the probability density.
        
        Instead of carrying out an explicit maximization, we monitor the distribution of the real and imaginary parts of the drift directly. Importantly, this approach does not render gauge cooling \cite{Seiler:2012wz, Dong:2020mtk} or other techniques based on complexified gauge symmetry irrelevant. Our analysis remains within the confines of unitary gauge symmetry, $\Omega_{x} \in SU(N)$, whereas gauge cooling operates within the larger complexified gauge space, partially fixing the non-unitary degrees of freedom.
        
        Adapted for lattice gauge theories, the correctness criterion for CL asserts that the CL method yields valid results if the probability density of the drift magnitude,
        \begin{align}
            \label{eq:drift_density_gauge}
            p_K(u, \theta) := \int \mathcal{D}U \, \delta(N_K[{U}]-u) P({U};\theta),
        \end{align}
        decays at least exponentially, where $P({U};\theta)$ is the probability density function characterizing the CL process for links. In this expression, the integration is carried out over configurations in the complexified gauge group $\mathrm{SL}(N, \mathbb{C})$, with $\mathcal{D}U$ denoting the corresponding Haar measure. This quantity can be directly evaluated using configurations sampled from CL simulations, providing a practical and mathematically rigorous method for validating correctness, especially in regimes or for observables inaccessible by other techniques.
        
        We emphasize, however, that the derivation of this criterion assumes full analyticity of the theory. Specifically, that the action and observables are entire functions and that all integrals converge absolutely. In cases where the action has singularities or branch cuts, i.e., other theories than the considered pure gauge theory, the criterion may not be applicable and a careful reassessment would become necessary.
        
        Even when all assumptions are satisfied, the numerical evaluation of the criterion is non-trivial. Effects such as Langevin time dependence, and the resolution and range of the drift density, can significantly influence the results. To mitigate these issues, we complement this criterion with additional diagnostics: we compare one-point correlation functions with Euclidean benchmarks and monitor the unitarity norm $F[U]$ as discussed below, which has empirically shown high sensitivity to instabilities while remaining bounded in stable simulations.

    \subsection{Correctness criterion with kernels} \label{sec:coc_kernels}

        The proof of the correctness criterion is based on the time evolution of the real probability density associated with the CL process, which is governed by the Fokker-Planck equation. To establish a correspondence with complex weights, we consider a complexified Fokker-Planck equation that governs the evolution of the complex weight function $\rho(\theta)$, satisfying $\rho(\theta \to \infty) \propto e^{-S}$. 
        A kernel $\Gamma$ does not alter the stationary complex weight, but it does affect both its CL evolution as well as the evolution and the associated real probability density of the CL process. Therefore, the argumentation for the criterion needs to take the kernel function into account.
    
        The correctness criterion can be formulated analogously for the case with a kernel. When analyzing the time evolution under the kernel substitution, it becomes evident that the definition of the drift magnitude must be appropriately modified,%
        \footnote{For simplicity, we assume a diagonal kernel structure; however, a corresponding definition of the drift magnitude also exists for more general kernels.}
        \begin{align}
            N^\Gamma_K := \max_{x, \mu} \sum_a \left( \left| \mathrm{Re}\, \Gamma_{x,\mu} K^a_{x,\mu} \right| + \left| \mathrm{Im} \,\Gamma_{x,\mu}K^a_{x,\mu} \right| \right).
        \end{align}
        We restrict our attention to kernel functions that guarantee convergence of the relevant integrals and allow for integration by parts. These conditions are implicitly assumed to hold in the non-kernelled case. It is crucial to choose the kernel carefully, as it can impact the foundational assumptions underlying the CL method. In this work, we focus on polynomial kernels, whose analytic structure in the complex plane provides useful constraints for the design and optimization of the method.
        
        Analogous to the case without a kernel, the correctness criterion now asserts that CL converges correctly if the density,
        \begin{align}
            p^\Gamma_K(u, \theta) := \int \mathcal{D}U\, \delta\left(N^\Gamma_K[U] - u\right)\, P_{\Gamma}(U; \theta),
        \end{align}
        exhibits at least exponential decay in $u$. Here, $P_{\Gamma}$ denotes the real probability density associated with the CL process in the presence of the kernel function $\Gamma$.
        
        It is important to emphasize that the introduction of a kernel function generally modifies the real probability distribution $P_{\Gamma}$ and may allow the correctness criterion to be satisfied in cases where it fails without a kernel. However, introducing a kernel requires caution: zeros and singularities of $\Gamma$ in the complex plane can lead to significant complications. Specifically, zeros of $\Gamma$ can stall the dynamics, while singularities may destabilize the solution of the CL equation. To prevent such issues, we impose the following conditions on $\Gamma$,
        \begin{align}
            \begin{split}
                &\int \mathcal{D}U \, P_{\Gamma}(U;\theta) \, \delta\left( \left| \Gamma(U) \right| \right) = 0, \\
                &\int \mathcal{D}U \, P_{\Gamma}(U;\theta) \, \delta\left( \frac{1}{\left| \Gamma(U) \right|} \right) = 0,
            \end{split}
        \end{align}
        for all Langevin times $\theta \geq 0$. These conditions ensure that the sets of zeros and singularities of the kernel function have measure zero with respect to the probability distribution of the process. As a consequence, they are almost surely not encountered and do not contribute. Nonetheless, numerical instabilities may still arise when the process approaches the vicinity of a singularity. In such cases, it may be necessary to stabilize the simulation using adaptive time-step control techniques \cite{Flower:1986hv, Aarts:2009dg}.
        
        It is important to note that the condition discussed above concerns the deterministic drift term as evaluated with respect to the probability distribution of the CL process. However, focusing solely on the modification of the drift through the introduction of a kernel function is not sufficient to ensure a faster decay of the drift distribution or to design an effective kernel. In practice, we find that the corresponding alteration of the diffusion term, affecting the probability density $P$ via the Fokker-Planck evolution, can counteract the intended effect. In some instances, this leads to a process that probes deeper regions of the complexified configuration space. Consequently, the design of kernel functions, especially for systems with many degrees of freedom, is a highly non-trivial task. Nonetheless, we observe that even simple kernel choices can be remarkably effective in practice. In particular, their application to $\mathrm{SU}(N)$ lattice Yang–Mills theory in four dimensions yields promising results, as we will explore in the following sections.
    
    \subsection{Further remarks on the validity of CL}
        
        The validity of the CL method in recovering a real representation of the path integral weight function relies on certain correctness criteria, none of which are unique. For instance, an alternative \textit{a posteriori} condition, derived in~\cite{Scherzer:2018hid}, is based on integration by parts applied to the evolution of the real probability density. A key drawback of this approach is that the boundary terms must be evaluated on an observable-by-observable basis. Since the primary objective is to compute non-local observables, verifying these terms becomes computationally intensive. In contrast, monitoring the decay behavior of the drift magnitude throughout the simulation offers a more practical diagnostic, as it does not require repeated checks across different observables. Consequently, we adopt this strategy as our primary diagnostic tool in numerical simulations.
    
        Moreover, these correctness criteria are typically applied retrospectively; that is, they can only be evaluated after the simulations have been performed. This presents a fundamental limitation: they offer little guidance on how to proceed when the conditions for correctness are not satisfied. The central challenge of the CL method is thus twofold. First, the conditions that lead to convergence failure are still not fully understood. Second, there is no systematic or constructive framework to ensure reliable convergence in concrete applications. Addressing these shortcomings requires further research, not only to refine diagnostic criteria but also to develop practical methods for adapting or correcting the CL approach in order to broaden its range of applicability.
    
        A promising route toward resolving both the diagnostic and constructive challenges of the CL method lies in the strategic use of kernel functions. Kernels modify the stochastic dynamics by reshaping both the drift and diffusion components of the CL evolution, thereby exerting control over how the complexified configuration space is explored. This flexibility enables one to suppress excursions into regions that dominate incorrect convergence and to guide the process toward domains that are physically relevant. In particular, kernels can be designed to enhance the decay of the drift distribution, thereby satisfying the correctness criterion \cite{Nagata:2016vkn} more robustly, as showcased, for instance, in \cite{Boguslavski:2022dee}. Moreover, by influencing the sampling profile, they offer a potential means of aligning the dynamics more closely with the dominant Lefschetz thimbles of the theory \cite{Aarts:2014nxa, Aarts:2013fpa, Okamoto:1988ru, Alvestad:2022abf, Hayata:2015lzj}. While the precise relationship between kernel design and thimble geometry remains to be fully understood, our results indicate that even simple kernel choices can significantly improve stability and convergence. Thus, kernels provide a practical and adaptable mechanism not only for diagnosing pathological behavior but also for systematically improving the reliability of CL simulations in complex quantum field theories.
    
\section{Field-independent phase kernels}
\label{sec:field-indep_kernels}

    In this section, we explore the potential of contour-dependent but field-independent phase kernels to improve the stability of CL simulations in the presence of a real-time sign problem. While non-real-valued field-independent kernels%
    \footnote{We note that the anisotropic kernel in Eq.~\eqref{eq:aniso_kernel} falls outside of this class of kernel functions.} 
    have already been shown to enhance convergence in low-dimensional models \cite{Okano:1991tz, Alvestad:2022abf, Alvestad:2023jgl}, their applicability to more realistic field theories is limited by the increased complexity of the dynamics and the presence of multiple degrees of freedom. Nevertheless, under certain symmetry considerations, it is possible to construct kernel functions that are independent of the field variables, yet retain stabilizing features. We begin by motivating the construction in the context of an analytically tractable zero-dimensional toy model, and then proceed to examine how symmetries inherent to lattice gauge theories can guide the formulation of more practical field-independent kernels for higher-dimensional systems.

    \subsection{Motivation via zero-dimensional models}

        \subsubsection{Model with a single interaction term}
        
        Consider a simple model with a single degree of freedom, $x \in \mathbb{R}$, governed by the action
        \begin{equation}
            S(x) = \beta x^n,
        \end{equation}
        where $\beta \in \mathbb{C}$ is a complex coupling constant leading to a numerical sign problem. Depending on the value of the coupling, the CL method is known to suffer from wrong convergence for this model. This can be circumvented by a simple constant kernel \cite{Okano:1991tz}. The corresponding CL equation, incorporating a constant complex kernel%
        \footnote{In the literature, the term ``kernel'' typically refers to the square of the function $\gamma$, i.e., to $\Gamma = \gamma^2$. In this work, we will refer to holomorphic $\gamma$ as kernel as well. Then the factorizability of $\Gamma$ is guaranteed without the need of further constraints.} 
        $\gamma$ and real-valued Gaussian noise, is given by
        \begin{align}
            \partial_{\theta} z(\theta) = -n \beta\, \gamma^2\, z(\theta)^{n-1} + \gamma \,\eta(\theta),
        \end{align}
        where the drift has been defined as $K = -\partial_z S$. 
        Here, $z = x + i y \in \mathbb{C}$ is the complexified degree of freedom as a function of the Langevin time $\theta$, the action has been analytically continued, and $\eta$ denotes the standard real noise term. We consider constant kernels in terms of complex phases $\gamma = e^{i\phi} \in \mathbb{C}$ since the magnitude of the kernel would effectively only rescale the Langevin time step without circumventing wrong convergence for zero-dimensional models.%
        \footnote{For the anisotropic kernel discussed in Eq.~\eqref{eq:aniso_kernel}, the situation is different. It uses the fact that gauge fields are Lorentz vectors and rescales the Langevin time step of different field components separately.}
        
        We motivate the choice of the phase angle $\phi$ of the kernel by considering a cold start for the CL simulations, taking $z(\theta = 0) = 0$ and initializing the process with an arbitrary noise draw $\eta(0) = r_0 \in \mathbb{R}$. After one update step with finite step size $\varepsilon$, the process evolves to 
        \begin{equation}
        \label{eq:model_firststep}
        z(\varepsilon) = \sqrt{\varepsilon}\, r_0 e^{i\phi}.
        \end{equation}
        Proceeding with another Langevin step with a second independent noise draw $\eta(\varepsilon) = r_1 \in \mathbb{R}$, we obtain
        \begin{align}
        \label{eq:model_secondstep}
        &z(2\varepsilon) \\
        &= z(\varepsilon) - n \varepsilon^{(n+1)/2} \beta_r e^{i\psi} r_0^{n-1} e^{i(n-1)\phi} e^{2i\phi} + \sqrt{\varepsilon}\, r_1 e^{i\phi}, \nonumber 
        \end{align}
        where the coupling has been expressed in polar form as $\beta = \beta_r e^{i\psi}$.
        
        To constrain the Langevin process along a fixed straight line in the complex plane, we require the phase of the process to remain constant throughout the Langevin evolution. Since the phase of the kernel $\phi$ modulates both drift and noise terms, and the noise is sampled randomly, we must ensure that all terms in the update have identical phases. Requiring this in the above expression leads to the condition 
        \begin{align}
            \psi + 2\phi + (n - 1)\phi = \phi,
        \end{align}
        which simplifies to 
        \begin{align}
        \label{eq:phase_kernel_cond_model}
            \phi = -\psi/n.
        \end{align}
        This condition ensures that $z(\theta) = r(\theta) e^{i\phi}$ with $r(\theta)\in\mathbb{R}$ along the entire Langevin trajectory, beyond the first steps in \eqref{eq:model_firststep} and \eqref{eq:model_secondstep}. This is easy to see from the CL equation for $r(\theta)$
        \begin{align}
            \partial_\theta r(\theta) = - n \beta_r r^{n-1}(\theta) + \eta(\theta), \quad r(0) = 0,
        \end{align}
        which is entirely real-valued and free of the original sign problem.
        However, we emphasize that this construction only works for models with a single degree of freedom and a single coupling parameter.

        \subsubsection{Model with multiple terms}
        
        For a more complicated model, we consider the anharmonic oscillator governed by the action 
        \begin{align}
            S(x) = \frac{\beta}{2} x^2 + \frac{\lambda}{4} x^4,
        \end{align}
        with $\beta = \beta_r e^{i\psi} \in \mathbb{C}$ and $\lambda \in \mathbb{R}$. In this case, the second Langevin step becomes
        \begin{align}
        &z(2\varepsilon)
        = z(\varepsilon) - \varepsilon \gamma^2 \left.\partial_z S\right|_{z=z(0)}
        + \sqrt{\varepsilon} \gamma \eta(\varepsilon) \\
        &= z(\varepsilon) \! - \! \varepsilon e^{2i\phi} \left( \beta_r e^{i\psi} \varepsilon^{1/2} r_0 e^{i\phi} \! + \! \lambda \varepsilon^{3/2} r_0^3 e^{3i\phi} \right) \! +\!  \sqrt{\varepsilon}\, r_1 e^{i\phi}. \nonumber
        \end{align}
        To confine the CL trajectory in the complex plane with a constant phase, in this case, multiple conflicting conditions would need to be satisfied
        \begin{align}
            \psi + 3\phi = 5\phi, \quad \psi + 3\phi = \phi, \quad 5\phi = \phi,
        \end{align}
        which are simultaneously valid only if $\psi = \phi = 0$. This corresponds to the trivial case where $\beta \in \mathbb{R}$ and $\gamma = 1$, i.e.\ a sign-problem-free model and neutral kernel.
        
        Therefore, we conclude that there is no general choice of the constant kernel that confines the Langevin process to a line in the complex plane for models involving multiple terms with different parameters and orders. The same can be shown when we introduce more degrees of freedom. Hence, a constant kernel cannot entirely avoid the sign problem and potentially the wrong convergence of CL.

        \subsubsection{Arguments for the existence of an ``optimal'' phase}
        
        Although a kernel function specified by a simple phase does not generally solve the convergence problems of CL dynamics, it is natural to ask whether there exist phase choices that at least improve the behavior. A useful indicator in this context is whether the kernel restricts the stochastic process to a straight line in the complexified configuration space, ideally aligned with a dominant integration contour. Such restriction can lead to correct sampling from the path integral.

        Consider the case $\beta = \beta_r e^{i\psi} \in \mathbb{C}$ and $\lambda = 0$. According to Eq.~\eqref{eq:phase_kernel_cond_model},
        choosing the kernel phase $\gamma = e^{-i\psi/2}$ ensures that the CL process evolves along a fixed line in the complex plane. Similarly, for $\beta = 0$ and $\lambda = r_{\lambda} e^{i\kappa} \in \mathbb{C}$, the choice $\gamma = e^{-i\kappa/4}$ leads to the same behavior. In both cases, the dynamics remain effectively one-dimensional and avoid excursions into directions that can typically cause incorrect convergence.
        
        However, when both $\beta$ and $\lambda$ are non-zero, this strategy breaks down. The reason lies in the structure of the relevant Lefschetz thimbles, which are the steepest descent paths originating from the stationary points of the action. 
        While for single-parameter cases the relevant thimble that contributes to the deformed path integral often coincides with a straight path, the general case involves multiple thimbles that are typically curved \cite{Hayata:2015lzj, Boguslavski:2024yto}. As a result, no constant kernel phase can keep the dynamics restricted to the relevant regions. The stochastic process then explores regions beyond the relevant thimble, making it harder to assess whether the conditions for correct convergence are still met.
        
        However, it is plausible that we can interpolate between the optimal kernels of the limiting cases to find a phase that improves convergence. This interpolation provides a practical way to explore improved configurations and explicitly test the correctness criterion.

        We can further motivate the existence of a stabilizing phase kernel by applying It\^o's lemma to the transformed process $u(z) = e^{i\phi_u} z$. This leads to the stochastic differential equation
        \begin{align}
        \label{eq:modelb_kernelled}
            \partial_\theta u(\theta) = -e^{i\phi_u} \gamma^2 \partial_z S\vert_{z=z(u(\theta))} + e^{i\phi_u} \gamma \eta(\theta).
        \end{align}     
        Choosing $\phi_u = -\phi$ for a phase kernel of the form $\gamma = e^{i\phi}$ simplifies the equation to
        \begin{align}
            \partial_\theta u(\theta) = -\partial_u \tilde S\vert_{u=u(\theta)}+ \eta(\theta),
        \end{align}
        where we introduced the effective action $\tilde S(u):= S(z(u))$ governing the stochastic process of the complexified variable $u$. This transformation allows us to identify an effective action for the kernelled CL process,
        \begin{align} \label{eq:model_eff_action}
            \tilde{S}(u) = \frac{\gamma^2 \beta}{2} u^2 + \frac{\gamma^4 \lambda}{4} u^4,
        \end{align}
        in which the complex phases of the original parameters $\beta$ and $\lambda$ are modified through the kernel. By tuning the kernel phase $\phi$, we effectively rotate $\beta$ and $\lambda$ in the complex plane by different angles, which may improve the convergence properties of the Langevin process, as it is known that the correct convergence of CL depends on the phases of these parameters \cite{Aarts:2013uza}.
        
        An \emph{a priori} method for determining the optimal kernel phase angle $\phi$ is currently unknown. Nevertheless, a simple scan over $\phi$ is computationally inexpensive and can already lead to improved behavior. Moreover, in higher-dimensional field-theoretic models, model-specific symmetries can often guide or constrain the choice of kernel, as we will discuss in the next section.

    \subsection{Phase kernel for real-time Yang-Mills theory} \label{sec:sym_phase_kernel}

        Although it has not yet been demonstrated how to choose a constant-phase kernel for systems with multiple degrees of freedom that confines the CL process to a slice of configuration space, we argued in the previous subsection that introducing such a kernel can nonetheless enhance stability. This stabilization arises because the kernel modifies the dynamics, effectively resulting in a different model that exhibits the same expectation values.

        We now turn to the application of the phase kernel to real-time Yang-Mills theory on complex time contours in 3+1 dimensions. 
        In Sec.~\ref{sec:update_indep_kernel} we introduced the discretized CL equation for field-independent kernels that are diagonal in all indices. 
        Here, we will first consider the discretized version of the general form of the kernel \eqref{eq:Gamma_gamma}, $\gamma^{ab}_{x x',\mu\nu}$, and then constrain its form based on physical assumptions.
        
        Determining an ``optimal'' kernel is generally difficult for two main reasons. Firstly, the notion of optimality itself is ambiguous: the CL dynamics may involve a non-diagonal diffusion matrix that lacks analytical control, and it is typically non-trivial to identify a well-defined effective action as for the model \eqref{eq:model_eff_action} discussed above. Even if such an action could be constructed, it remains unclear whether it would lead to improved convergence in practice. Secondly, assuming an optimization scheme were available, the dimensionality of the kernel matrix increases rapidly with the lattice volume and number of degrees of freedom, rendering direct optimization computationally prohibitive.%
        \footnote{We note that in lower-dimensional scalar field theories, machine-learned kernels have shown promising results~\cite{Alvestad:2023jgl}, though extending such approaches to higher-dimensional gauge theories poses significant computational challenges.} 
        
        We therefore assess the potential of a field-independent phase kernel by leveraging manifest symmetries of the theory and adopting several simplifying assumptions for the kernel structure:

        \begin{itemize}
            \item \textbf{Spatial homogeneity:} The system is isotropic and homogeneous in the spatial directions, which the kernel should inherit.
        
            \item \textbf{Locality:} The original drift $K$ is quasi-local, i.e., $K(x)$ depends only on fields in a small neighborhood.
            We therefore consider kernel functions that yield a local kernel-modified drift $K^\Gamma$ and diffusion term.
        
            \item \textbf{Euclidean neutrality:} The Euclidean contour branches $\mathcal{C}_\mathrm{E}$ in Figure~\ref{fig:sk_contour} neither induce instabilities nor a sign problem; accordingly, the kernel should remain neutral along those segments of the lattice.
        
            \item \textbf{Time reflection symmetry:} The forward and backward parts of the regularized (tilted) Schwinger-Keldysh time contour $\mathcal{C}_\pm$, depicted in Figure~\ref{fig:sk_contour}, have been assigned phase factors of opposite argument.
            Hence, the phases of the kernel along these contour parts should be conjugated with respect to each other.%
            \footnote{We have empirically checked that this choice gives the best results by combining the phases for forward/backward branches independently.}
        
            \item \textbf{Directional diagonality:} For simplicity, we consider no mixing between different directions in the kernel, leading to a diagonal structure in the directional degrees of freedom.
        
            \item \textbf{Magnitude governed by anisotropy:} We assume that the anisotropic kernel \eqref{eq:aniso_kernel} governs the magnitude of the kernel function, up to an overall constant that can be absorbed in the Langevin time step $\epsilon$. Therefore, we restrict the free parameters to the complex phases.

            \item \textbf{Color neutrality:} 
            Since no color component is enhanced, we assume independence of color indices. This ensures that the kernel does not break the gauge covariant structure of the CL equations.
        \end{itemize}

        Taking these considerations into account, the field-independent kernel takes the form
        \begin{align} \label{eq:const_phase_kernel}
            \gamma^{ab}_{xx',\mu\nu} = \delta^{ab}\delta_{\mu\nu} \delta_{xx'} \gamma_{x,\mu},
        \end{align}
        as was already assumed in Eq.~\eqref{eq:CL_latt_gamma_indep}. 
        The argument $x$ is written in full spacetime notation for notational simplicity, even though $\gamma_{x,\mu}$ effectively depends only on the contour time. The function $\gamma_{x,\mu}$ is then defined as
        \begin{align} \label{eq:const_phase_kernel_param}
            \gamma_{x,\mu} :=
            \begin{cases}
                e^{\pm i\phi_t/2}/\xi, & \text{if} ~~ \mu = 0,~ x \in \mathscr{C}^\mathrm{rect}_{\pm}, \\
                e^{\pm i\phi_s/2},       & \text{if} ~~ \mu \in \{1,2,3\},~ x \in \mathscr{C}^\mathrm{rect}_{\pm}, \\
                1/\xi,                 & \text{if} ~~ \mu = 0,~ x \in \mathscr{C}^\mathrm{rect}_E, \\
                1,                       & \text{if} ~~ \mu \in \{1,2,3\},~ x \in \mathscr{C}^\mathrm{rect}_E.
            \end{cases}
        \end{align}
        The time path $\mathscr{C}^\mathrm{rect}$ refers to the discretized rectangular contour shown in Figure \ref{fig:sk_contour}, with $\mathscr{C}^\mathrm{rect}_{\pm}$ denoting the forward and backward branches tilted into the negative imaginary direction, and $\mathscr{C}^\mathrm{rect}_E$ representing the two vertical segments at $\mathrm{Re}(t) = 0$ and $\mathrm{Re}(t) = t_\mathrm{max}$. While we focus on this specific contour, the kernel construction can, in principle, be straightforwardly generalized to alternative contour choices.
        
        The phase angles $\phi_t, \phi_s \in [0, 2\pi)$ are therefore the only free parameters that require optimization under the above assumptions, and the magnitude is dictated by the anisotropic kernel with anisotropy $\xi = a_s / |a_0|$. This makes it feasible to search for optimal parameters with respect to stability by scanning the unit circle with sufficient granularity. Moreover, this approach may reveal connections between model parameters, such as the tilt angle, coupling, or anisotropy $\xi$, and the optimal phase angles $\phi_t, \phi_s$.
        
        We note that the assumptions leading to this simple field-independent kernel are unlikely to yield a truly optimal kernel in the sense of stabilizing simulations for arbitrary model parameters. Nonetheless, the study of this kernel is important: It serves as a baseline for more advanced algorithmic optimizations, including field-dependent kernels (see Sec.~\ref{sec:field-dep_kernels}), allows us to assess the stabilizing potential of kernel modifications, and even with limited improvements, can extend our access to the real-time dynamics of quantum fields, a domain that remains largely unexplored.
        
    \subsection{Improved stability with the phase kernel} \label{sec:phase_kernel_results}

        \begin{figure*}[tb]
            \centering
            \includegraphics[width=0.8\textwidth]{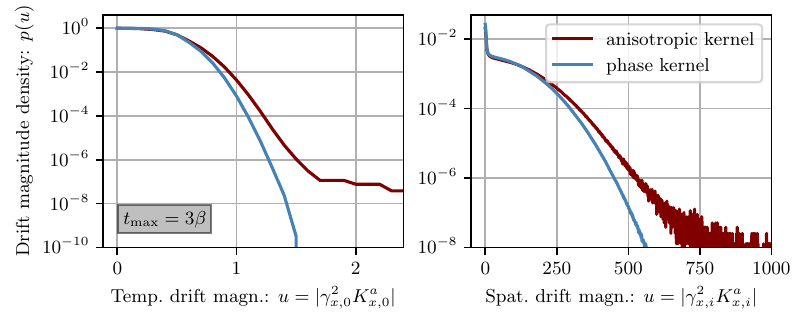}
            \caption{Densities of the temporal (\emph{left}) and spatial (\emph{right}) drift magnitude $p(u)$ for CL simulations on the rectangular SK contour with $t_\mathrm{max}=3\beta$ and $\tan(\alpha)=1/100$. The dark red lines represent simulations with only the anisotropic kernel, while the blue curves include the contour-dependent phase kernel from Eq.~\eqref{eq:const_phase_kernel} with $\phi_t = \pi/4$ and $\phi_s = 0 $. The latter setting leads to exponential decay of the drift magnitude, satisfying the correctness criterion.}
            \label{fig:pu_phase}
        \end{figure*}
        
        We assess the impact of the contour-dependent phase kernel defined in Eqs.~\eqref{eq:const_phase_kernel} and \eqref{eq:const_phase_kernel_param} on the stability of CL simulations. We will use the same model parameters as in Ref.~\cite{Boguslavski:2023unu}: a coupling constant $g = 0.5$ and the inverse temperature $\beta = 1 $. 
        For all tests, we employ the regularized rectangular SK contour depicted in Figure \ref{fig:sk_contour} with a real-time extent of $t_{\mathrm{max}} = 3\beta$ discretized on a $N_t\times N_s^3=224\times 16^3$ lattice and the values for $|a_0|$, $a_s$, and $\alpha$ specified in Sec.~\ref{sec:CL_lattice_gauge}.
        We focus here on evaluating the stabilizing effect of the kernel modification while leaving its application to real-time observables to Ref.~\cite{companion:2026}. In addition to the kernel modification, we apply gauge cooling \cite{Seiler:2012wz} using a single cooling step per CL update with a cooling stepsize of $\alpha_\mathrm{gc} = 0.05 $. The Langevin evolution is performed using a higher-order integration scheme with a fixed stepsize $\epsilon = 10^{-4} $, as detailed in Appendix~\ref{app:higherorder_sde}.

        In our previous work \cite{Boguslavski:2023unu}, we used an SK contour discretization with isosceles-tilted real-time paths, where we reached $t_{\mathrm{max}} = 1.5\beta$ for the same theory and discretization parameters. The rectangular SK contour depicted in Figure \ref{fig:sk_contour} in combination with the anisotropic kernel and gauge cooling permits stable simulations up to $t_{\mathrm{max}} = 2\beta$, which serves as a useful benchmark for the present study of stabilizing kernels. In particular, we significantly extend the accessible real-time extent here to $t_{\mathrm{max}} = 3\beta$ at comparable tilt angles to exhaust the stabilizing effect of the anisotropic kernel for the considered anisotropy.

        The stabilizing effect of the phase kernel is presented in Fig.~\ref{fig:pu_phase}, where we show the drift magnitude distributions $p(u)$ for the temporal (left) and spatial (right) components of the CL process. As discussed in Sec.~\ref{sec:coc_gamma}, this quantity is central to verify the validity of the CL simulations according to the criterion of correctness. 
        For simulations using only the anisotropic kernel \eqref{eq:aniso_kernel} on the rectangular contour (red curves), we observe a power-law tail in $p(u)$ indicating a violation of the correctness criterion. 
        In contrast, when the contour-dependent phase kernel with phases $\phi_t = \pi/4$ and $\phi_s = 0$ is applied in combination with the anisotropic kernel,%
        \footnote{Due to severe numerical instabilities in the absence of the anisotropic kernel, it is not feasible to determine the drift density or observables reliably without it. We therefore adopt the simulation with the anisotropic kernel as our benchmark.}
        the drift distributions decay exponentially in both components. This behavior satisfies the correctness criterion and restores the formal consistency of the CL approach under these conditions.%
        \footnote{
        We note that while the absolute magnitude of the drift is less relevant than its asymptotic decay for the criterion of correctness, the temporal drift component is significantly smaller than its spatial counterpart due to the anisotropic kernel rescaling it by a factor $1/\xi^2 = 1/1024 $.}
        
        \begin{figure}[tb]
            \centering
            \includegraphics[width=0.48\textwidth]{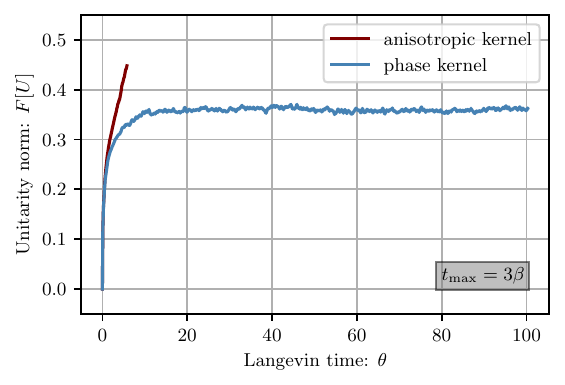}
            \caption{Unitarity norm as a function of Langevin time $\theta$ for CL simulations on the rectangular SK contour with $t_\mathrm{max}=3\beta$ and $\tan(\alpha)=1/100$. The dark red curve corresponds to simulations with the anisotropic kernel only and exhibits rapid divergence. The blue curve shows simulations with the phase kernel from Eq.~\eqref{eq:const_phase_kernel} using $\phi_t = \pi/4 $, which maintain a small and stable unitarity norm throughout the evolution.}
            \label{fig:un_phase}
        \end{figure}

        \begin{table}[t]
            \footnotesize
            \centering
            \caption{
                Real and imaginary parts of planar spatial Wilson loops $W^s_{k \times k}$ of order $k $, comparing results from CL simulations on the SK contour ($t_\mathrm{max} = 3 \beta$, $g=0.5$) denoted by $\langle\cdot \rangle_{\mathrm{SK}}$ using the phase kernel with $\phi_t = \pi/4$ and $\phi_s = 0$ to benchmark values from a Euclidean contour denoted by $\langle\cdot \rangle_{\mathrm{E}}$. The standard errors, shown in parentheses, are bootstrap-estimated and account for autocorrelations. The agreement within uncertainties confirms the correctness of the CL method with the phase kernel.
            }\label{tab:phase}
            \begin{tabular}{c || p{1.8cm} p{1.8cm} | p{1.8cm} p{1.8cm}}
                \toprule
                ~~$k$~~ & $\mathrm{Re} \langle W^s_{k\times k} \rangle_{\mathrm{E}}$ & $\mathrm{Re} \langle W^s_{k\times k} \rangle_{\mathrm{SK}}$ & $\mathrm{Im} \langle W^s_{k\times k} \rangle_{\mathrm{E}}$ & $\mathrm{Im} \langle W^s_{k\times k} \rangle_{\mathrm{SK}}$\\
                \midrule
                 $1$ & $1.95456(9)$ & $1.9545(1)$ & $0.0$ & $\phantom{-}0.0000(1)$\\
                 $2$ & $1.8753(5)$ & $1.8749(3)$ & $0.0$ &  $\phantom{-}0.0001(3)$\\
                 $3$ & $1.786(1)$ & $1.7857(5)$ & $0.0$ & $-0.0001(5)$\\
                \bottomrule      
            \end{tabular}
        \end{table}
        
        As an independent cross-check, the stabilizing effect of the phase kernel is further evidenced by the behavior of the unitarity norm 
        \begin{align} 
        \label{eq:unitarity_norm}
            F[U] = \frac{1}{4N_t N_s^3}\sum_{x,\mu} \textrm{Tr} \left[ \left( U_{x,\mu} U_{x,\mu}^\dagger - \mathbbm{1} \right)^2 \right].
        \end{align}
        It is shown in Figure \ref{fig:un_phase} as a function of the Langevin time for both kernels on the considered real-time extent. When using the anisotropic kernel alone, the norm exhibits rapid growth after a short transient period, signaling a breakdown of the simulation due to uncontrolled excursions into the non-unitary direction of the complexified configuration space. In contrast, the inclusion of the contour-dependent phase kernel keeps the unitarity norm bounded and sufficiently small. 
        
        To further validate the correctness of the CL simulations with the phase kernel, we check time translation invariance that is inherent to thermal field theory. 
        In particular, we compare gauge-invariant observables, which do not depend on the time contour, against known reference results. Table~\ref{tab:phase} shows the real and imaginary parts of planar spatial Wilson loops of size $k \times k $, $\langle W^s_{k\times k} \rangle$, computed using the phase kernel with the previously stated simulation parameters. The results from simulations on the SK contour are consistent with those obtained on a sign-problem-free Euclidean contour (Eucl), and all discrepancies lie within statistical uncertainties. These results confirm that the simulations not only exhibit stable Langevin dynamics but also yield quantitatively correct expectation values for physical observables, thereby strengthening the case for the practical utility of the contour-dependent phase kernel.

        We note that for simplicity, the spatial phase parameter in Eq.~\eqref{eq:const_phase_kernel} was fixed to $\phi_s = 0$ to reduce the dimensionality of the parameter scan. While coarse scans did not reveal further improvements by varying $\phi_s $, a more systematic exploration may uncover beneficial parameter choices. We also emphasize that the choice $\phi_t = \pi/4$ is not unique: we observed correct convergence within the range $\phi_t \in [\pi/8, 3\pi/8] $.

        These results demonstrate that even a simple kernel modification, contour-dependent but field-independent, can substantially improve stability and correct convergence in CL simulations of real-time gauge theories. Compared to the baseline setup on the rectangular SK contour, the use of the phase kernel allows us to extend the maximal real-time contour by 50\% from $2\beta$ to $3\beta$.
        
        The success of this kernel motivates further investigation of more general, possibly field-dependent, kernel constructions. Such extensions could offer enhanced control over the Langevin dynamics by structurally changing the underlying stochastic differential equation, and are discussed in the following section.

\section{Field-dependent kernels}
\label{sec:field-dep_kernels}

    Building upon the insights gained from constant-phase kernels, we now turn to the construction and analysis of \emph{field-dependent} kernels for CL simulations of real-time SU($N$) Yang-Mills theory. These kernels offer increased flexibility by allowing local, configuration-dependent modifications of both the drift and diffusion terms, enabling more refined control over stability and convergence. However, this generalization also introduces new challenges, such as the appearance of functional derivatives in the update equations and the potential of worsening rather than improving the convergence properties of the problem.

    In this section, we construct a class of holomorphic, gauge-invariant, and local field-dependent kernels motivated by both theoretical considerations and empirical observations. A key feature of our approach is that the kernel depends only on a small number of parameters and on the field configuration via traced plaquette variables, making it relatively simple to tune. As a result, the optimization of kernel parameters does not require computationally intensive procedures and can be performed via low-dimensional parameter scans.
    
    Similar to the strategy of the previous section, we begin by outlining the general structural requirements that such kernels should satisfy, including analyticity, locality, and gauge covariance. We then introduce a specific kernel structure based on local plaquette observables and contour-dependent parameters and assess its effect on numerical stability and correct convergence. Alternative constructions, including polynomial, non-polynomial, and exponential forms, are also discussed at the end of this section.

    \subsection{Introduction of field-dependent kernels}

        For clarity, we reiterate the discrete update step of the CL process for the link fields, Eq.~\eqref{eq:link_update_gamma},
        \begin{align} \label{eq:link_update_gamma_fkernel}
            \begin{split}
                U_{x,\mu}(\theta   +   \epsilon) =\, &e^{i t^a \left(\epsilon K^{\Gamma, a}_{x,\mu}[U]+ \sqrt{\epsilon} \, \gamma_{x,\mu}[U]\, \eta^a_{x,\mu}\right)} \, U_{x,\mu}(\theta),\\
                it^a K^{\Gamma,a}_{x,\mu} =\,& \gamma_{x,\mu}^2 \, it^a K^a_{x,\mu} + 2 \gamma_{x,\mu} \, \gamma'_{x,\mu}.
            \end{split}
        \end{align}
        The diagonal structure of the kernel \eqref{eq:Gamma_diag_gamma} that reduces to $\gamma_{x,\mu}[U]$ is motivated by the same arguments as laid out in Sec.~\ref{sec:sym_phase_kernel}. 
        
        In contrast to field-independent kernels, the field-dependent kernels introduce an additional functional derivative, $i t^a D^a_{x,\mu}\Gamma_{x,\mu}[U] = 2\gamma_{x,\mu} \gamma_{x,\mu}'$, and the diffusion term becomes effectively field dependent, $\gamma_{x,\mu}[U] \, \eta_{x,\mu}$, where $\gamma'_{x,\mu}$ denotes the group derivative of the kernel function with respect to $U_{x,\mu}$,
        \begin{align}
            \gamma'_{x,\mu} = i t^a D^a_{x,\mu} \gamma_{x,\mu}.
        \end{align}
        
        In the following, we motivate our choice of the kernel function $\gamma_{x,\mu}$ based on empirical observations, while discussing the broader functional requirements imposed on field-dependent kernels in addition to the symmetry considerations outlined in Section \ref{sec:sym_phase_kernel}. Since this class of modifications is significantly larger than the constant case, we impose the following additional structural constraints to guide the construction:
        
        \begin{itemize}
            \item \textbf{Holomorphicity:} The kernel function $\gamma_{x,\mu}$ must be holomorphic. This is essential to preserve the analyticity assumptions underlying the CL method, its justification via the complex contour deformation, and the assumptions underlying the criterion of correct convergence of \cite{Nagata:2016vkn} as discussed in Sec.~\ref{sec:coc_gamma}.
            
            \item \textbf{Local neutrality:} The kernel should be parametrized such that it remains locally connected to the identity (i.e., the neutral kernel $\gamma_{x,\mu} = 1$). This ensures that it can be ``switched off'' along Euclidean branches of the contour $\mathcal{C}_E$, where no sign problem or instabilities occur. If the kernel remained active in these regions, its complex-valued structure could amplify non-unitary deviations and destabilize the dynamics.
            
            \item \textbf{Locality of the drift:} We require the derivative $\gamma'_{x,\mu}$ to be local. While the physical interpretation of the kernel is still subtle, non-local kernels may induce volume-scaling artifacts and propagate instabilities across distant lattice sites. This undermines the efficacy of gauge cooling and other local stabilization techniques.
            
            \item \textbf{Gauge covariance:} The kernel must preserve gauge invariance in its functional form, and its derivative should transform covariantly. This ensures that the kernel-modified drift respects gauge symmetry and that the gauge cooling procedure remains valid \cite{Nagata:2015uga}.
        \end{itemize}
        
        In analogy to the field-independent phase kernel, we parametrize the new kernel using a minimal set of tunable parameters to systematically improve stability. In the next subsection, we will motivate the specific kernel structure used in this work and evaluate its impact on the CL convergence. We emphasize that the choice of field-dependent kernel is inherently model-specific: both the kernel derivative and the modified diffusion term must be evaluated in the context of the original drift to assess their effect. The resulting dynamics of the Langevin process are significantly more complex than in the field-independent case. The associated diffusion matrix of the corresponding real Fokker–Planck equation becomes non-diagonal and nontrivial to analyze. Therefore, we limit ourselves here to exploring empirically motivated kernel structures, testing their efficacy across representative parameter regimes, and comparing the results with those obtained previously.
        
        Finally, we emphasize that the introduction of field-dependent diffusion modifies the convergence properties of standard numerical integrators. The Euler–Maruyama scheme converges only at first order $\mathcal{O}(\epsilon)$. Therefore, as discussed in Sec.~\ref{sec:higher_order_maintext} and detailed in Appendix~\ref{app:higherorder_sde}, we employ a Runge–Kutta-type scheme for SU($N$) lattice gauge theories in analogy to the improved step. This scheme retains second-order convergence $\mathcal{O}(\epsilon^2)$ even for field-dependent noise and allows for a comparison with the results from the anisotropic kernel and the phase kernel.
        
    \subsection{Contour- and field-dependent kernels} \label{sec:const_lin_kernel}

        In the following, we investigate the impact of field-dependent kernels of the form
        \begin{align} \label{eq:fd_kernel}
            &\Gamma_{x,\mu} = 1 + \! \sum_{y \in V_{x,\mu}}  \sum_{\nu \neq \mu} m_{y,\mu\nu} \, (h_{y,\mu\nu} \!  + \! h_{y,\nu\mu})
            \\ &h_{y,\mu\nu} = \mathrm{Tr}[U_{y,\mu\nu}],
        \end{align}
        where $V_{x,\mu}$ denotes a local neighborhood volume of the site $x$, and $m_{y,\mu\nu} \in \mathbb{C}$ are direction- and spacetime-dependent kernel parameters. This form of the kernel is a natural extension of the phase kernel introduced in Section \ref{sec:sym_phase_kernel} as it can be interpreted as a linear polynomial in the traced plaquette with complex coefficients, while the phase kernel is a constant polynomial. We did not include the phase kernel directly to allow a separate analysis of the impact of the field dependence and allow a tuning of a lower-dimensional parameter space. The rationale behind this construction is outlined below.

        The volume $V_{x,\mu}$ is chosen to include all plaquette variables that depend on the link $U_{x,\mu}$ when taking the functional derivative of $\Gamma_{x,\mu}$ with respect to $U_{x,\mu}$. This minimal set is a natural choice, as it reflects the structure of the original drift term, which arises from an action integrated over the entire lattice and thereby depends on all links. 
        We note that although supersets of this minimal volume yield the same derivative, they introduce a potentially unwanted non-locality into the update step. Importantly, the kernel function is extensive with respect to $V_{x,\mu}$, i.e., its magnitude and fluctuation increase with the size of the neighborhood. While one could normalize the kernel by a constant volume factor, doing so reduces the strength of its derivative term and weakens its stabilizing effect. On the other hand, since the thermal expectation value of the trace of the plaquette is real, averaging over the lattice introduces controlled fluctuations around this stable value. This improves control over the kernel and reduces the propagation of non-unitary behavior in the complexified manifold, especially in regions where the gauge links are already non-unitary. Without such control, these distortions can be amplified by the kernel and destabilize the simulation.
        
        In this work, we test a local kernel structure defined by the minimal neighborhood $V_{x,\mu} := \{ x \} \cup \prod_{\nu \neq \mu} \{ x-\hat\nu\}$ for all $x \in \Omega$ and directions $\mu \in \{0,1,\dots,d\}$, where $\Omega$ denotes the simulated lattice volume. 
        We have also tested larger neighborhood volumes, but the benefits were inconclusive. In particular, we observed a significant increase in autocorrelation length for larger volumes, likely due to the increased non-locality in the update dynamics. A more detailed study of this effect would require precise estimation of the autocorrelation scale for each case and is reserved for future work.
        
        The choice of the parameters $m_{\mu\nu}(x)$ follows from symmetry considerations, similarly to the phase kernel in Eq.~\eqref{eq:const_phase_kernel_param}. Specifically, we choose
        \begin{align}
        \label{eq:fd_kernel_param_first}
            m_{x,ii} = &m_{x,i0} = 0, \quad m_{x,0i} := m_{x,0} \in \mathbb{C}, \\ &m_{x,ij} = m_{x,ji} = m_{x,i} \in \mathbb{C}
        \end{align}
        for all $i,j \in \{1,\dots, d\}$ with $i \neq j$ and define the functional form of $m_{x,\mu}$ as
        \begin{align} \label{eq:fd_kernel_param}
            m_{x,\mu} =
            \begin{cases}
                r_t e^{\pm i\varphi_t},  & \text{if} ~~ \mu = 0,\quad x \in \mathscr{C}^\mathrm{rect}_{\pm}, \\
                r_s e^{\pm i\varphi_s},  & \text{if} ~~ \mu \in \{1,2,3\},\quad x \in \mathscr{C}^\mathrm{rect}_{\pm}, \\
                0,                   & \text{otherwise.}
            \end{cases}
        \end{align}
        A detailed motivation for these choices was provided prior to Eq.~\eqref{eq:const_phase_kernel}. The discretized rectangular contour $\mathscr{C}^\mathrm{rect}$ in Fig.~\ref{fig:sk_contour} consists of the forward and backward real-time branches $\mathscr{C}^\mathrm{rect}_{\pm}$, which are tilted into the lower half of the complex time plane, and the Euclidean segments $\mathscr{C}^\mathrm{rect}_E$ located at $\mathrm{Re}(t) = 0$ and $\mathrm{Re}(t) = t_\mathrm{max}$. Although our analysis is based on this particular contour, the kernel formulation can be readily extended to other contour choices.
        
        In addition to the phase angles $\varphi_t$ and $\varphi_s$, the magnitudes $r_t$ and $r_s$ serve as tunable parameters that weight the plaquette sum relative to the unity term. This flexibility is crucial on anisotropic lattices, where temporal and spatial plaquettes typically exhibit different magnitudes. The sign of the phase angles again encodes symmetry across the forward and backward real-time contour branches, while the kernel is set to be neutral on the Euclidean parts of the contour. Moreover, by setting $m_{x,i0} = 0$, we ensure that temporal plaquettes enter only into the temporal update scheme, and spatial plaquettes only into the spatial update. This separation of directional contributions has been empirically found to enhance numerical stability.
        
        The derivative of the kernel enters the drift term of the update equation for the link variable in Eq.~\eqref{eq:link_update_gamma_fkernel} via
        \begin{align} 
            &\Gamma'_{x,\mu} = 2 \gamma_{x,\mu} \gamma'_{x,\mu} = i t^a D^a_{x,\mu} \Gamma_{x,\mu} \nonumber \\
            &~= -t^a \sum_{\nu \neq \mu} m_{x,\mu\nu}\, \mathrm{Tr}[t^a (U_{x,\mu \nu} - U_{x,\nu \mu}] \\
            &~\quad -t^a \sum_{\nu \neq \mu} m_{x-\hat\nu,\mu\nu}\, \mathrm{Tr}[t^a (U_{x,\mu(-\nu)} - U_{x,(-\nu)\mu}]. \nonumber
        \end{align}
        We note that, in principle, one could attempt to match the kernel structure to the Wilson action or its real or imaginary components. However, our numerical experiments indicate that such constructions tend to increase instabilities rather than mitigate them.
        
        We combine the field-dependent kernel from above with the anisotropic kernel and the rectangular regularized SK contour to check their joint influence on the correctness of the results. The anisotropy $\xi$ introduces an overall rescaling factor in the temporal direction, further modulating the kernel structure and its stabilizing effect.

    \subsection{Numerical results}
    
        \begin{figure*}[tb]
            \centering
            \includegraphics[width=0.8\textwidth]{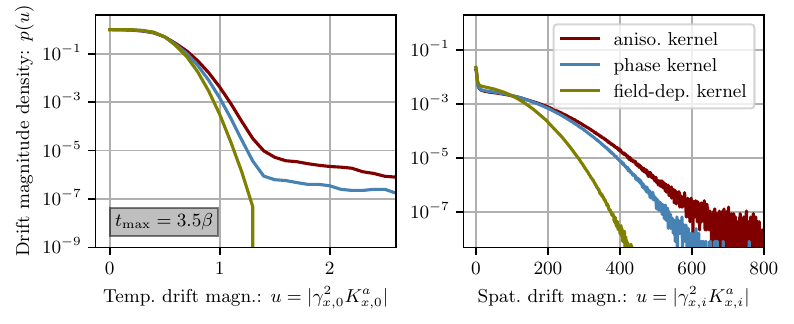}
            \caption{Densities of the temporal (\emph{left}) and spatial (\emph{right}) drift magnitude $p(u)$ for CL simulations on the rectangular SK contour with $t_\mathrm{max}=3.5\beta$ and $\tan(\alpha)=1/100$. The dark red lines represent simulations with only the anisotropic kernel, the blue curves include the contour-dependent phase kernel from Eq.~\eqref{eq:const_phase_kernel} with $\phi_t = \pi/4$ and $\phi_s = 0$ and the green line shows the results for the field dependent kernel with $\varphi_t = 51\pi/100, r_t=1/8$ and $\varphi_s = r_s=0 $. The latter setting leads to exponential decay of the drift magnitude, satisfying the correctness criterion.}
            \label{fig:pu_fd}
        \end{figure*}
        We assess the stability improvements of the field-dependent kernel by comparing it to the phase kernel and the anisotropic kernel for the same model and simulation parameters as summarized in Sections \ref{sec:CL_lattice_gauge} and \ref{sec:phase_kernel_results}. To exploit the limitation of these reference kernels, we simulate a maximal real-time extent to $t_{\mathrm{max}}=3.5\beta$. As in the previous section, we employ the fixed tilt angle $\tan(\alpha)=1/100$ and an anisotropy of $\xi=32$, which leads to a lattice size of $N_t\times N_s^3=256\times 16^3$. 
        In analogy to before, we combine both the phase and the field-dependent kernels with the anisotropic kernel. 
        
        We performed a parameter scan of the phase kernel to tune the phase parameters in Eq.~\eqref{eq:const_phase_kernel_param} and found that the same parameter set as for $t_{\mathrm{max}} = 3\beta$ achieves the best stability. The results without any kernel could not be directly studied, as the severe numerical instabilities of the CL simulation disallow the evaluation of drift magnitude histograms or benchmarking observables, such that at least the anisotropic kernel has to be included.

        In Figure \ref{fig:pu_fd}, we show the drift magnitude density $p(u)$ with $t_{\mathrm{max}} = 3.5\beta$ for the temporal (left) and spatial (right) components of the kernelled CL drift for the three different kernel types. We observe that the anisotropic kernel and the phase kernel exhibit a power law decay, violating the criterion of correctness for CL and therefore yield wrong results. We also observe a numerical breakdown of the simulation for sufficiently long simulation times. We emphasize that the phase kernel does lead to a faster decay in $u$ compared to only the anisotropic kernel, but not to the necessary exponential decrease of the convergence criterion, and hence it is not sufficient for this setting. In contrast, the drift magnitude density with the field-dependent kernel \eqref{eq:fd_kernel} with parameters $\varphi_t = 51\pi/100, r_t=1/8$ and $\varphi_s = r_s=0$ entering the contour-dependent kernel coefficients in Eq.~\eqref{eq:fd_kernel_param}, yields an exponential decay of the drift magnitude density. Hence, the simulations for the field-dependent kernel satisfy the criterion of correctness and thus provide correct field configurations.
        
        \begin{figure}[tb]
            \centering
            \includegraphics[width=0.48\textwidth]{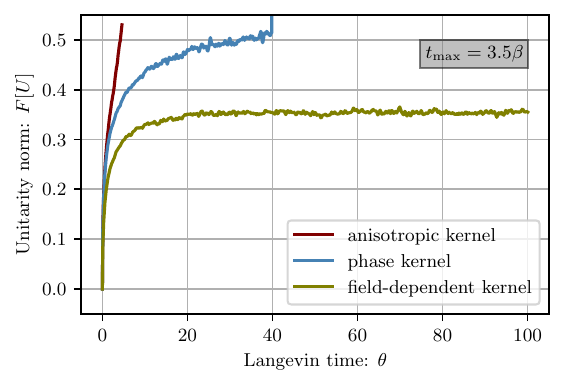}
            \caption{Unitarity norm as a function of Langevin time $\theta$ for CL simulations on the rectangular SK contour with $t_\mathrm{max}=3.5\beta$ for the same kernels and parameter choices as in Fig.~\ref{fig:pu_fd}. 
            While the pure ansitropic and the (additional) phase kernel lead to divergent curves, simulations with the field-dependent kernel in addition to the anisotropic kernel maintain a small and stable unitarity norm throughout the evolution.}
            \label{fig:un_fd}
        \end{figure}
        
        These observations can also be visualized by the unitarity norm \eqref{eq:unitarity_norm}, which we show in Fig.~\ref{fig:un_fd} as a function of Langevin time $\theta$. For the anisotropic kernel, the norm steeply grows due to a strongly non-unitary evolution, leading to a rapid breakdown of the simulation. The phase kernel mitigates these fast excursions of the Langevin trajectory into the non-unitary direction of the configuration space. However, it does not stabilize the simulations indefinitely, as observed by the gradual growth of the non-unitarity, leading to an eventual breakdown of the simulations around $\theta=40$. The field-dependent kernel avoids the problem of a rising non-unitarity, marked by a constant unitarity norm, also for long Langevin times. We conclude that for all practical purposes, our CL simulations with the field-dependent kernel satisfy the criterion of correctness and lead to stable Langevin trajectories.

        As in the previous section, we cross-check this conclusion by comparing specific observables to benchmark results. Table \ref{tab:fd} presents the real and imaginary parts of planar spatial Wilson loops of size $k \times k$, calculated using the field-dependent kernel. Within statistical uncertainties, the real-time results agree with those obtained on a Euclidean contour, which is free of a sign problem. This agreement confirms, for the example of Wilson loops, that the simulations not only maintain stable Langevin dynamics but also lead to correctly distributed lattice configurations. This reinforces the effectiveness and practical relevance of the field-dependent kernel.

        We emphasize that the tuned angle $\varphi_t=51\pi/100=:\pi/2+\phi$ entering Eq.~\eqref{eq:fd_kernel_param} appears to be connected with the tilt angle of the contour $\alpha$ entering the complex temporal lattice spacing $a_0=e^{i\phi}/32$ in the present discretization of the SK-contour. We empirically tested and confirmed this connection by varying the tilt angle $\alpha$ of the contour and $\varphi_t$ to find the optimal relation 
        \begin{align}
            \varphi_t \approx \pi[1/2+\tan(\alpha)].
        \end{align}
        However, this empirical relation is only found in the absence of the spatial component of the field-dependent kernel, but it reduces the parameter search to a scan of the magnitude $r_t$.
        
        \begin{table}[t]
            \footnotesize
            \centering
            \caption{
                Real and imaginary parts of planar spatial Wilson loops $W^s_{k \times k}$ of order $k $, comparing results from real-time CL simulations ($t_\mathrm{max} = 3.5\beta$, $g=0.5$) denoted by $\langle\cdot \rangle_{\mathrm{SK}}$ using the field-dependent kernel with $\varphi_t = 51\pi/100, r_t=1/8$ and $\varphi_s = r_s=0$ to benchmark values from a Euclidean contour denoted by $\langle\cdot \rangle_{\mathrm{E}}$. The errors shown in parentheses are bootstrap-estimated and account for autocorrelations. The agreement within uncertainties confirms the correctness of the CL method with the field-dependent kernel.
            }\label{tab:fd}
            \begin{tabular}{c || p{1.8cm} p{1.8cm} | p{1.8cm} p{1.8cm}}
                \toprule
                ~~$k$~~ & $\mathrm{Re} \langle W^s_{k\times k} \rangle_{\mathrm{E}}$ & $\mathrm{Re} \langle W^s_{k\times k} \rangle_{\mathrm{SK}}$ & $\mathrm{Im} \langle W^s_{k\times k} \rangle_{\mathrm{E}}$ & $\mathrm{Im} \langle W^s_{k\times k} \rangle_{\mathrm{SK}}$\\
                \midrule
                 $1$ & $1.95456(9)$ & $1.9546(1)$ & $0.0$ & $-0.0001(1)$\\
                 $2$ & $1.8753(5)$ & $1.8760(6)$ & $0.0$ &  $-0.0003(3)$\\
                 $3$ & $1.786(1)$ & $1.791(5)$ & $0.0$ & $-0.0004(5)$\\
                \bottomrule      
            \end{tabular}
        \end{table}
        The parameter tuning was only conducted for the temporal parameters $\varphi_t$ and $r_t$ in Eq.~\eqref{eq:fd_kernel_param} and $\varphi_s=r_s=0$ to reduce the dimensionality of the parameter scan. We note, however, that including also the spatial part of the kernel potentially leads to further improved convergence, which requires extension in future work. Also, a combination of the phase kernel and the field-dependent kernel may lead to further improvements, but needs further investigation.
        
    \subsection{Other field-dependent kernels}

        In the previous subsections, we proposed and motivated a specific class of field-dependent kernels designed to enhance the stability of CL dynamics in real-time SU($N$) Yang–Mills theory. That construction was facilitated by model symmetries and engineered to enable systematic parameter scans. In this section, we briefly report on a non-exhaustive set of alternative field-dependent kernels we have tested, which either did not lead to a significant stabilization improvement or performed worse than our primary choice.
        
        \subsubsection{Variants of the field-dependent kernel construction}
        
            As discussed in Section~\ref{sec:const_lin_kernel}, the kernel can be modified by altering both the local lattice volume $V_x$ and the parameter structure. We conducted preliminary tests with larger volumes and relaxed the constraint $m_{i0} = 0$ imposed in Eq.~\eqref{eq:fd_kernel_param_first}. However, neither change yielded consistent improvements in stability. In some cases, changes in autocorrelation length made the effects difficult to interpret conclusively. We also experimented with lifting additional symmetry constraints to allow off-diagonal kernel components and more flexible parameter choices. While this may open the door to more powerful kernel constructions, such flexibility requires more sophisticated optimization strategies, potentially involving machine learning or adaptive schemes.
            
            The kernel in Eq.~\eqref{eq:fd_kernel} is affine in the plaquette trace variables. We therefore tested generalized polynomial kernels of the form
            \begin{align}
                \tilde h_{y,\mu\nu} = \mathrm{Tr}[U_{y,\mu\nu}]^k, \qquad k = 2, 3, 4.
            \end{align}
            However, higher-order powers tended to amplify fluctuations in the kernel function, ultimately degrading stability. We attribute this to increased sensitivity to large excursions in field space, which are especially problematic near the boundaries of the complexified manifold.
            
            We further investigated non-polynomial kernel functions of the form
            \begin{align}
                \hat h_{y,\mu\nu} = f(\mathrm{Tr}[U_{y,\mu\nu}]),
            \end{align}
            where, for example, $f(\cdot) = \arctan(\cdot/N)$ and $f(\cdot) = 1/(1+(\cdot/N)^2)$ were studied. The rationale behind this choice is that $\arctan(\cdot)$ is bounded on the real axis, thus is well-behaved near the real manifold. However, in complex directions, this function becomes unbounded and exhibits poles on the imaginary axis. These features break fundamental analyticity assumptions required for the validity of the CL method, and our tests confirmed that such kernels introduce instabilities in practice.
            
        \subsubsection{Exponential kernel functions}
        
            An intriguing exponential kernel form was proposed in \cite{Soderberg:1987pd} for a far simpler model. Adapted to real-time Yang-Mills theory, it is given by
            \begin{align}
                \Gamma_{x,\mu} = \exp\left\{ \sum_{y \in V_{x,\mu}} \sum_{\nu \neq \mu} m_{y,\mu\nu} \, \mathrm{Tr}[U_{y,\mu\nu}+U_{y,\nu\mu}] \right\},
            \end{align}
            where the key idea is to choose the parameters $m_{\mu\nu}(x) = -i{\rho_{x,\mu\nu}/g^2}$ from the Wilson action \eqref{eq:wilson_action} such that the derivative of the exponent cancels the original drift. This yields a Langevin process governed entirely by the diffusion term, effectively eliminating the drift. It was argued in Ref.~\cite{Soderberg:1987pd} that for real actions in compact domains, such a construction can converge correctly. However, for non-compact domains, the unbounded exponential growth of the kernel leads to divergence.
            
            We tested this kernel in Euclidean lattice simulations for SU($2$) Yang–Mills theory. To avoid extensivity and non-locality issues, we restricted the kernel to local lattice volumes $V_x$ rather than integrating over the whole lattice. Under these conditions, we found that the process converged correctly, albeit with significantly longer thermalization times. In contrast, when applying the exponential kernel to real-time simulations on a regularized SK contour, we observed strong instabilities. These are likely due to uncontrolled fluctuations in the kernel function, which grow exponentially in the complexified field space and introduce severe numerical challenges.
            
            It is worth noting that these simulations were performed using the standard Euler–Maruyama integration scheme. We have not yet tested this kernel using higher-order solvers, as for the field-dependent kernel discussed in the previous section, which may help mitigate some of the observed instabilities.

\section{Conclusion} 
\label{sec:conclusion}
    
    In conjunction with its forthcoming paper that focuses on unequal-time correlations \cite{companion:2026}, this study achieves the first successful application of field-dependent kernel modifications to complex Langevin simulations of a full quantum field theory in four spacetime dimensions.
    
    We demonstrated that such kernels, constructed with symmetry-guided design principles, lead to significant improvements in numerical stability and correct convergence across extended real-time contours. Both contour-dependent phase kernels and configuration-sensitive field-dependent kernels were shown to restore the correctness of CL dynamics in regimes previously dominated by uncontrolled excursions. This led to the significant extension of the real-time contour from previously $t_{\textrm{max}} = 1.5 \beta$ in \cite{Boguslavski:2023unu} and $t_{\textrm{max}} = 2 \beta$ on a rectangular time contour to $t_{\textrm{max}} = 3 \beta$ and $t_{\textrm{max}} = 3.5 \beta$ for the field-independent phase kernel and the field-dependent kernel we introduced. 
    In \cite{companion:2026} we will employ this framework to compute unequal-time correlation functions for unprecedentedly long real-time extents, bringing us closer to physically relevant predictions. 
    
    A key insight from this work is the crucial role of symmetry considerations in guiding the kernel construction. Constraints such as spatial homogeneity, time-reversal symmetry, and neutrality along the Euclidean contour were instrumental in reducing the complexity of the parameter space and identifying effective kernel structures. Future work may build on this strategy by incorporating additional theoretical input, such as known thermal expectation values or constraints from Dyson–Schwinger equations, to guide the kernel design in a more principled manner. These approaches hold great potential for advancing the applicability of CL to real-time dynamics in non-perturbative quantum field theories.

    As an essential validation step for the correct convergence of the kernelled complex Langevin process, we extended the established criterion of correctness based on the distribution of the drift magnitude \cite{Nagata:2016vkn} to kernelled CL in lattice gauge theories. Applying this extended criterion, we confirmed that the kernels introduced here indeed ensure correct convergence.

    The use of stochastic higher-order update schemes in this study was motivated primarily by the need for comparability with earlier results obtained using the anisotropic kernel in combination with the ``improved step'' scheme, which achieves second-order convergence in Langevin time. For field-dependent kernels, both the Euler–Maruyama scheme and the ``improved step'' are not sufficiently accurate. We therefore formulated a higher-order update scheme compatible with field-dependent kernels and implemented it in the context of lattice gauge theory.

    Taken together, these developments represent an important step toward first-principles calculations of real-time correlation functions in non-Abelian gauge theories at unprecedented temporal extents. They bring us closer to bridging the long-standing gap between lattice gauge theory and the real-time physics essential for understanding the dynamics of strongly interacting matter.
    
    The promising results obtained here open several directions for further investigation. A natural next step involves implementing more systematic parameter scans and optimization procedures to identify and cross-correlate kernel parameters across a broader range of model parameters. Moreover, combining the phase kernel and the field-dependent kernel may offer additional stability benefits and potentially allow for further extension of the real-time domain. Finally, the framework developed here paves the way for the computation of more sophisticated observables towards physically relevant predictions. Preliminary tests indicate that the presented kernel modifications may also stabilize simulations at larger bare couplings $g$, bringing real-time CL closer to the scaling regime. This, in turn, may open the possibility of performing continuum extrapolations in future studies.

\begin{acknowledgments}
    We would like to thank Gert Aarts, Jan Pawlowski, Alexander Rothkopf, and Denes Sexty for discussions. 
    This work is funded in whole or in part by the Austrian Science Fund (FWF) under the projects \mbox{[10.55776/P34455]}. 
    D.M.~acknowledges additional support from project P~34764. The computational results presented have been achieved in part using the Vienna Scientific Cluster (VSC). We acknowledge the VSC also for awarding the project access to the LEONARDO supercomputer, owned by the EuroHPC Joint Undertaking, hosted by CINECA (Italy) and the LEONARDO consortium.
\end{acknowledgments}

\appendix

\section{Higher order solver} \label{app:higherorder_sde}
    
    In this part of the appendix, we discuss numerical solvers for kernelled CL equations. To achieve convergence beyond leading order in the discretization of Langevin time, the commonly used `improved’ step becomes inadequate for field-dependent kernels. Instead, more refined numerical methods are needed. We illustrate the key aspects using a zero-dimensional stochastic process and comment at the end on the natural extension to lattice quantum field theories.

    \subsection{The Euler–Maruyama scheme}
        
        The standard method for simulating stochastic differential equations (SDEs) of the form
        \begin{align}
        \label{eq:sde_general_0D}
        dx(\theta) = K(x(\theta))\, d\theta + \sigma(x(\theta))\, dW(\theta),
        \end{align}
        is the Euler-Maruyama (EM) scheme. Here, $x(\theta)$ is a stochastic process, $K(x)$ denotes the drift term, $\sigma(x)$ the diffusion term, and $W(\theta)$ is a standard Wiener process (Brownian motion). 
        
        Discretizing Langevin time as $\theta_0, \theta_1, \ldots, \theta_n$ with step sizes $\epsilon_i = \theta_{i+1} - \theta_i $, the numerical update is given by
        \begin{align}
        x_{i+1} = x_i + K(x_i)\, \epsilon_i + \sigma(x_i)\, \Delta W_i,
        \end{align}
        where $x_i = x(\theta_i) $, and $\Delta W_i = W(\theta_{i+1}) - W(\theta_i) \sim \mathcal{N}(0, \epsilon_i)$ denotes a Gaussian-distributed Wiener increment with zero mean and variance $\epsilon_i $.
        
        The EM scheme achieves weak convergence of order one, which means that for any sufficiently smooth observable function $f$ and a fixed final Langevin time $\Theta $,
        \begin{align}
        \left| \mathbb{E}[f(X(\Theta))] - \mathbb{E}[f(x_{\epsilon}(\Theta))] \right| = O(\epsilon),
        \end{align}
        where $X(\Theta)$ denotes the exact solution of the stochastic differential equation at time $\Theta $, and $x_{\epsilon}(\Theta)$ is the numerical approximation obtained using a discretization with step size $\epsilon $. The notation $\mathbb{E}[\cdot]$ refers to the expectation value over realizations of the stochastic process. Since CL simulations aim to compute expectation values of observables, weak convergence is indeed the relevant criterion for assessing numerical accuracy.
        
        Due to its simplicity and low computational cost, EM remains the standard choice in many CL applications. It extends naturally to field-theoretic models, provided that the discretized dynamics respect the algebraic and geometric structure of the field space. However, the increasing relevance of kernelled CL dynamics highlights a major limitation of the scheme: for state-dependent diffusion terms $\sigma = \sigma(x(\theta))$, the strong convergence rate drops from order one to order one-half.

        This loss in convergence order introduces three significant complications. First, the complexification required in CL formulations typically leads to unbounded drift and diffusion terms. This amplifies discretization errors and necessitates smaller step sizes, increasing computational cost while still being limited by the low accuracy of EM. Second, when comparing different stabilization strategies, such as field-independent versus field-dependent kernels, numerical consistency is essential. Any observed differences must be disentangled from discretization artifacts, which is only possible if all tests use the same scheme with sufficient convergence control. Third, the criterion of correct convergence our study uses, which is based on the decay of the drift magnitude, derives from a finite-time expansion of the probability density. If the convergence rate is too low, deviations arising from large effective step sizes can mimic or obscure pathological behavior. In such cases, higher-order methods are not desirable to reliably diagnose failures of the method’s assumptions.

    \subsection{The stochastic Runge-Kutta scheme}
    
        We discuss a Runge–Kutta scheme for SDEs designed to achieve weak second-order convergence by introducing intermediate evaluations $v$, $v^+$, and $v^-$ within each time step
        \begin{align}
        v &= x_i + K(x_i)\, \Delta t + \sigma(x_i)\, \Delta W_i, \\
        v^+ &= x_i + K(x_i)\, \Delta t + \sigma(x_i)\, \sqrt{\Delta t}, \\
        v^- &= x_i + K(x_i)\, \Delta t - \sigma(x_i)\, \sqrt{\Delta t}.
        \end{align}
        The update rule is then given by
        \begin{align}
        x_{i+1} = x_i
        &+ \frac{1}{2}\big(K(v) + K(x_i)\big)\, \Delta t \nonumber \\
        &+ \frac{1}{4}\big(\sigma(v^+) + \sigma(v^-) + 2\sigma(x_i)\big)\, \Delta W_i \nonumber \\
        &+ \frac{1}{4} \frac{\sigma(v^+) - \sigma(v^-)}{\sqrt{\Delta t}} \big(\Delta W_i^2 - \Delta t\big),
        \end{align}
        which is notably more involved than the EM step. In particular, its practical implementation requires additional care, especially in memory-intensive settings such as lattice field theory simulations.
        
        In the context of CL without kernels, this scheme is commonly referred to as the improved step, originally introduced in \cite{Ukawa:1985hr} for lattice gauge theory with constant diffusion terms. There, it effectively amounts to averaging the drift between the current state $x_i$ and the intermediate evaluation $u$, which was shown to improve numerical stability. In Ref.\ \cite{Boguslavski:2023unu}, we confirm that this improvement persists in our simulations and leads to more robust dynamics under similar conditions.
        
        However, when CL kernels are introduced, see Section \ref{sec:kernel_CL_update}, the situation becomes more intricate. The dynamics now involve non-trivial, state-dependent diffusion terms, and the simplified improved step is no longer applicable. To retain consistency, the update must incorporate diffusion-dependent contributions arising from the kernel. Since the Wiener increment is Gaussian and thus unbounded, unaccounted fluctuations in the diffusion term can lead to significant numerical instabilities. Averaging the diffusion over intermediate evaluations, as done in the second-order scheme, becomes essential in this setting. The structure of the method naturally allows for such averaging and offers a systematic approach to suppress discretization artifacts induced by state-dependent noise. Moreover, implementing the scheme with kernels enables a consistent framework to assess the stability improvements achieved through field-dependent kernels, in direct comparison with alternative stabilization methods.
    
    \subsection{Extension to lattice gauge theory}

        The stochastic Runge-Kutta procedure must be carefully adapted to manifolds to preserve the algebraic structure during the numerical evolution. This is naturally achieved in analogy with Runge-Kutta-Munthe-Kaas (RK-MK) methods for differentiable Lie groups \cite{munthe1998runge}, suitably generalized to stochastic differential equations. We describe the extension of this scheme to lattice gauge theories, starting from the continuous CL equation for gauge fields, given in Eq.~\eqref{eq:cle}. The fundamental degrees of freedom are the complexified gauge fields $ A_\mu(x) \in \mathfrak{sl}(N, \mathbb{C}) $.
        
        The scheme is constructed by first discretizing the CL equation at the level of the gauge fields. As in the zero-dimensional case, we define intermediate evaluations of drift and diffusion terms within each Langevin step. These are used to build a Lie algebra-valued increment $\Xi^a_{x,\mu}$ combining the contributions of drift and noise. The gauge link update is then defined via exponentiation
        \begin{align}
        U_{x,\mu}(\theta + \epsilon) = \exp\left( i t^a \, \Xi^a_{x,\mu} \right) U_{x,\mu}(\theta),
        \end{align}
        where the expression for $\Xi^a_{x,\mu}$ is constructed to achieve weak second-order convergence. A heuristic derivation of this structure follows from the Baker-Campbell-Hausdorff formula. Its practical implementation mirrors the logic of Eqs.~(C.38) and (C.39) in Ref.~\cite{Boguslavski:2022dee}, now generalized to include kernelled drift terms and state-dependent diffusion.
        
        This formulation preserves gauge covariance by construction and ensures the numerical evolution remains on the group manifold through the exponential map. It thus provides a consistent framework for higher-order CL simulations in lattice gauge theories, including those with non-trivial kernel structures. In practice, the intermediate link configurations $V, V^\pm$ are obtained by exponentiating suitable combinations of drift and noise applied to the current link variables, guaranteeing that all updates remain within the complexified gauge group.

\section{Criterion of correctness} \label{app:coc}
    
    For completeness and better legibility, we first rederive in this appendix the criterion for the correctness of CL dynamics introduced in Ref.~\cite{Nagata:2016vkn} and then extend it to kernelled CL processes. This criterion provides a practical, though \emph{a posteriori}, method to assess whether the probability density obtained from the CL process correctly reproduces the complex path integral weight. 

    \subsection{Restating the argument}
    
    We start our discussion by motivating the complex counterpart of the real Fokker-Planck (FP) equation 
    \begin{align}\label{eq:fpe}
        &\frac{d P_{Z}}{d\theta}(x,y; \theta) \nonumber \\
        &= \sum_{i=x,y} \partial_i \left[ -K_i  P_{Z}(x,y; \theta) + N_i \partial_i P_{Z}(x,y; \theta)\right] \nonumber \\
        &=:L^{\top} P_{Z}(x,y; \theta),
    \end{align}
    which governs the evolution of the real probability density of the complex Langevin process for a zero-dimensional dynamical variable $Z(\theta) = X(\theta) +i Y(\theta)$ described by the associated CL equation. The transpose of the real FP operator acts on observable functions and is given by
    \begin{align}
        L := \sum_{i = x, y} \left( + K_i \partial_i + N_i \partial_i^2 \right).
    \end{align}
    This operator satisfies the relation
    \begin{align}
        \langle P , L \mathcal{O} \rangle = \langle L^\top P , \mathcal{O} \rangle,
    \end{align}
    for all observables $\mathcal{O} $, where we introduce the shorthand notation
    \begin{align}
        \langle f, g \rangle := \int dx \, dy \, f(x, y) \, g(x, y).
    \end{align}
    Note that this is not a true inner product, as the functions involved may be complex-valued, which violates positive definiteness.
    
    Since we restrict our analysis to holomorphic observables, we can exploit the Cauchy–Riemann equations, $i \partial_x \mathcal{O} = \partial_y \mathcal{O} $, to eliminate the explicit $y $-derivatives. Hence, the operator $L$ simplifies to the ``complexified'' form
    \begin{align}
        L_c := + K \, \partial_x + (N_x - N_y) \, \partial_x^2,
    \end{align}
    where $K$ is the complex drift evaluated along the real axis. This operator allows us to introduce another \emph{complex Fokker-Planck equation} describing the aforementioned evolution of a complex weight function.
    
    \subsubsection{Stationary solution}

        We aim to demonstrate that the limiting real probability density $P_{\mathrm{st}}$ correctly characterizes the complex weight function $\rho(x)$. For this purpose, we introduce a conjectured (complex) FP equation of the form
        \begin{align} \label{eq:cfpe}
            \frac{d\rho}{d\theta}(x; \theta) = L^\top_c \rho(x; \theta),
        \end{align}
        where the adjoint operator $L^\top_c$ is given by
        \begin{align}
            L^\top_c = \partial_x \left(-K + (N_x - N_y) \partial_x\right).
        \end{align}
        The condition $N_x = N_y + 1$ now appears naturally justified. Under this constraint, the complex weight $\rho(x) = \exp[-S(x)]$ with $K = -\partial_x S$ satisfies the stationary equation
        \begin{align}
            L^\top_c \rho(x) = 0,
        \end{align}
        and is thus a stationary solution of the complex FP equation. This provides a formal evolution equation for the complex weight with the desired stationary distribution. It remains to be shown that the CL process converges to this solution in the stationary limit, that is, whether the evolution of the real probability density $P_{Z}$ and the complex weight $\rho$ lead to identical expectation values.
        
        \subsubsection{Time evolution}
        
        We compare the time evolution of the complex weight function and the real probability density by imposing the initial condition
        \begin{align} \label{eq:ic}
            \rho(x; \theta) \, \delta(y) = P_{Z}(x, y; \theta) \quad \text{at} \quad \theta = \theta_0,
        \end{align}
        where $P_{Z}(\cdot; \theta_0)$ is determined by the initial conditions of the CL equation.\footnote{We implicitly assume that both the real and complex FP equations are universal in the sense that they evolve toward the same stationary solution, regardless of initial conditions.} For notational clarity, we define two separate functions corresponding to expectation values with respect to $P_{Z}(\cdot; \theta)$ and $\rho(\cdot; \theta) $, respectively,
        \begin{align}
            \Phi(\theta) &:= \int dx \, dy \, P_{Z}(x, y; \theta) \, \mathcal{O}(x+iy), \\
            \Psi(\theta) &:= \int dx \, \rho(x; \theta) \, \mathcal{O}(x).
        \end{align}
        
        Given the choice of initial conditions, it follows immediately that%
        \footnote{The initial condition is chosen for convenience but is not essential. Any initial condition satisfying $\Phi(\theta_0) = \Psi(\theta_0)$ is sufficient for the argument that follows, provided the observables are regular and the relevant integrals converge.}
        \begin{align}
            \Phi(\theta_0) &= \int dx \, dy \, P_{Z}(x, y; \theta_0) \, \mathcal{O}(x + i y) \\
            &= \int dx \, \rho(x; \theta_0) \, \mathcal{O}(x)  = \Psi(\theta_0).
        \end{align}
        The next step is to establish that this equality holds for all $\theta > \theta_0 $. Assuming the correct time evolution of the real probability density, the desired identity follows in the stationary limit, where $P_{\mathrm{st}} = \lim_{\theta \to \infty} P_{Z}(\cdot ; \theta)$,
        \begin{align}
            \int dx \, \rho(x) \, \mathcal{O}(x) = \int dx \, dy \, P_{\mathrm{st}}(x, y) \, \mathcal{O}(x + i y).
        \end{align}
        We emphasize that this line of reasoning represents the standard justification for the correctness of the CL method and has been shown to hold in a variety of models, though it often omits potentially important mathematical subtleties.

        In practical applications, we simulate the CL process numerically, and hence it is natural to consider finite-time evolutions. Following the approach of Nagata et al.\ in Ref.\ \cite{Nagata:2016vkn}, we expand the expectation value $\Phi(\theta)$ at $\theta\geq\theta_0$ to obtain
        \begin{align} \label{eq:tevolve}
            \Phi(\theta + \Delta \theta) = \sum_{n=0}^{\infty} \frac{1}{n!} \Delta \theta^n \, \frac{d^n \Phi}{d\theta^n}(\theta),
        \end{align}
        where the $n $-th derivative is obtained by iteratively applying the FP operator $L $,
        \begin{align}
            \frac{d^n \Phi}{d\theta^n}(\theta) = \int dx  dy \, P_{Z}(x, y; \theta) \, L^n \mathcal{O}(x + i y).
        \end{align}
        The validity of this expansion requires that all integrals on the right-hand side exist and that the series converges uniformly. Accordingly, we assume the existence of a finite radius of convergence $r_{\mathrm{conv}} > 0 $, such that the series in Eq.~\eqref{eq:tevolve} converges for all $\Delta \theta < r_{\mathrm{conv}}$ and $\theta > \theta_0 $.

        {\em Base case:} 
        Since we consider the initial condition in Eq.~\eqref{eq:ic}, we have $\Phi(\theta_0) = \Psi(\theta_0)$ for arbitrary observables $\mathcal{O} $. We can therefore write for $\theta=\theta_0$
        \begin{align} \label{eq:helpereq}
            \int dx \, \rho(x;\theta) L_c^n \mathcal{O}(x) = \int dx dy \, P_{Z}(x,y; \theta) L^n \mathcal{O}(x+iy),
        \end{align}
        for any $n \in \mathbb{N} $. This identity, valid at the initial time, forms the base case of an induction argument establishing its validity for $\theta > \theta_0 $.
        
        {\em Induction step:} We consider $\theta' = \theta + \Delta \theta $, where $\Delta \theta < r_{\mathrm{conv}} $ and where we assume that Eq.~\eqref{eq:helpereq} holds for a $\theta \geq \theta_0$ that may be later than the initial Langevin time. This allows us to sum both sides of Eq.~\eqref{eq:helpereq}, and by the uniform convergence assumed earlier, we may exchange the order of integration and summation. This yields
        \begin{align} \label{eq:exprhoP}
            \int dx\, &\rho(x;\theta) e^{\Delta \theta L_c} \mathcal{O}(x)\nonumber \\ 
            =&\int dx dy\, P_{Z}(x,y; \theta) e^{\Delta \theta L} \mathcal{O}(x+iy).
        \end{align}
        By differentiating both sides $n$ times with respect to $\Delta \theta $, we obtain
        \begin{align}
            \int dx\, &\rho(x;\theta) e^{\Delta \theta L_c} \left( L_c^n \mathcal{O}(x) \right) \nonumber\\
            =&\int dx dy\, P_{Z}(x,y; \theta) e^{\Delta \theta L} \left(L^n \mathcal{O}(x+iy)\right).
        \end{align}
        Applying the transposed operators $L_c^\top$ and $L^\top$ to the probability densities $\rho$ and $P $, respectively, we arrive at
        \begin{align}
            \int dx\, &\rho(x;\theta+\Delta \theta ) L_c^n \mathcal{O}(x)  \nonumber\\
            = &\int dx dy\, P_{Z}(x,y; \theta+\Delta \theta) L^n \mathcal{O}(x+iy),
        \end{align}
        which completes the induction step. Thus, Eq.~\eqref{eq:helpereq} holds for all $\theta \geq \theta_0 $.

        {\em Consequences:} 
        From the case $n = 0 $, it immediately follows that $\Phi(\theta + \Delta\theta) = \Psi(\theta + \Delta\theta) $, and thus $\Phi(\theta') = \Psi(\theta')$, for all $\theta' \geq \theta_0 $. We conclude that the entire evolution of both $\rho$ and $P$ yields identical expectation values for analytically continuable observable functions $\mathcal{O} $.
        
        This proof relies on two key assumptions, in addition to the regularity of the functions involved:
        \begin{enumerate}
            \item The initial conditions for the FP equations in Eqs.~\eqref{eq:fpe} and~\eqref{eq:cfpe} are chosen such that they yield identical expectation values at $\theta_0 $.
            \item The series expansion for the expectation value evolution in Eq.~\eqref{eq:tevolve} converges uniformly.
        \end{enumerate}
        We note that the expansion of the left-hand side of Eq.~\eqref{eq:exprhoP} is justified under the assumption that the time evolution is governed by the corresponding FP equations, provided the system and its weight function are well-defined and sufficiently regular. In addition to the regularity of the observables, we also restrict to observables that are absolutely integrable with respect to both $P_{Z}(\cdot; \theta)$ and $\rho(\cdot; \theta)$ for all $\theta $. The first condition is necessary to establish the induction base and poses no fundamental difficulty, as it merely constitutes a constructive step in the proof. The second assumption is discussed in the following paragraph.
    
    \subsubsection{Condition for correct probability evolution.}
    
        To guarantee the uniform convergence of the expansion in Eq.~\eqref{eq:tevolve}, it was shown in Ref.~\cite{Nagata:2016vkn} that the probability density of the drift magnitude,
        \begin{align} \label{eq:drift_density}
            p(u, \theta) =& \int dx \, dy \, \delta(u(x + i y) - u) \, P_{Z}(x, y; \theta), \\
            u(x + i y) :=& \vert K(x + i y) \vert,
        \end{align}
        with respect to $P_{Z}(\cdot; \theta)$, has to decay at least exponentially for all $\theta > \theta_0 $.
        
        In the expansion of Eq.~\eqref{eq:tevolve}, the dominant contributions arise from integrals of the form
        \begin{align}
            c_n := \,& \frac{1}{n!}\int dx \, dy \, u^n(x + i y) \, P_{Z}(x, y; \theta) \nonumber \\
            = \,& \frac{1}{n!}\int_0^\infty du \, u^n \, p(u, \theta),
        \end{align}
        and we denote the corresponding series of coefficients by $c_n $. 
        To establish that the drift distribution $p(u, \theta)$ must decay at least exponentially for all Langevin time $\theta $, we apply the root test and the Cauchy–Hadamard theorem to the coefficients $c_n $. This yields the radius of convergence of the series
        \begin{align}
            R(\theta)&:=\left( \limsup_{n \to \infty} |c_n|^{1/n} \right)^{-1} \nonumber \\
            &= \left( \limsup_{n \to \infty} \left| \frac{1}{n!} \int_0^\infty du\, u^n\, p(u, \theta) \right|^{1/n} \right)^{-1}.
        \end{align}
        Assuming the asymptotic form $p(u, \theta) \sim e^{-r(\theta) u}$, we find
        \begin{align}
            R(\theta) \propto \left( \lim_{n \to \infty} \left| \frac{1}{n!} \int_0^\infty du\, u^n\, e^{-r(\theta) u} \right|^{1/n} \right)^{-1} = r(\theta),
        \end{align}
        showing that the radius of convergence is given by the exponential decay rate $r(\theta) $. Hence, absolute convergence of the series, and therefore the correctness of the CL method, requires that $r(\theta)$ is strictly positive and bounded from below, $r(\theta) \geq r_{\mathrm{conv}} > 0$, for all $\theta $. This concludes the proof that an exponential decay of the drift distribution with a uniform lower bound on the decay rate is indeed required for reliable CL convergence.
        
        Finally, we note that the whole discussion about the criterion of correctness focused on the zero-dimensional setting. We emphasize that all steps can be naturally extended to higher-dimensional models and field theories with gauge symmetries. We note that for these adoptions, we have to change derivatives to functional derivatives and the integral measure to the corresponding group measure. Moreover, the drift magnitude function $u$ has to respect direction and the non-complexified gauge degrees of freedom. For SU($N$) Yang-Mills lattice gauge field theory, such an extension is discussed in Section \ref{sec:coc_lattice} with the probability density of the draft magnitude given explicitly in Eq.~\eqref{eq:drift_density_gauge}.

    \subsection{Incorporating kernel freedom} \label{app:coc_w_kernels}

        The proof of the correctness criterion relies on the time evolution of the real probability density associated with the stochastic process, which is governed by the FP equation in Eq.~\eqref{eq:fpe}. To establish a correspondence with complex weights, we introduce a complexified FP equation, given in Eq.~\eqref{eq:cfpe}, which governs the evolution of the complex weight function $\rho $. 
        
        This correspondence between the real and complex FP equations is not unique. In particular, it permits the introduction of a family of complex FP equations that admit the path integral weight $\rho$ as a stationary solution. One way to generate such generalizations is by modifying the CL equation $dZ_i(\theta) = K_i \, d\theta + \sigma_{ij} \, dW_j(\theta)$ through the inclusion of a holomorphic \emph{kernel function} $\gamma $, leading to
        \begin{align} \label{eq:gamma_cle}
            dZ_i(\theta) = K^\gamma_i \, d\theta + \sigma^\gamma_{ij} \, dW_j(\theta),
        \end{align}
        where $i = x, y$ refer to the real and imaginary parts of the complex stochastic process $Z = X + i Y =: Z_x + i Z_y $. The Wiener increments $dW_x$ and $dW_y$ are assumed to be independent.
        
        The kernel-modified drift and diffusion terms are given by
        \begin{align} \label{eq:kernelled_drift}
            K^\gamma &= K_x^{\gamma} + i K_y^{\gamma} = \gamma^2 K + 2 \gamma \gamma', \\
            \sigma^\gamma &= \begin{pmatrix}
                \sqrt{2 N_x} \, \gamma_x & -\sqrt{2 N_y} \, \gamma_y \\
                \sqrt{2 N_x} \, \gamma_y & \sqrt{2 N_y} \, \gamma_x 
            \end{pmatrix},
        \end{align}
        where $\gamma'$ denotes the complex derivative of the holomorphic kernel function $\gamma $.\footnote{We remark that in the presence of multiple degrees of freedom, the kernel function can modify the evolution of each component independently. In such a case, the squared term $\gamma^2$ in the kernelled drift is replaced by the matrix-valued expression $\Gamma_{\mu\nu} = \sum_\kappa \gamma_{\mu\kappa} \gamma_{\nu\kappa} $, where the indices $\mu, \nu$ label the degrees of freedom. This structure is often referred to in the literature as \emph{factorizable}, although it arises naturally when the kernel is constructed from its underlying factor components. A concrete realization of such kernels in a field-theoretic context with multiple degrees of freedom are introduced in Sections~\ref{sec:field-indep_kernels} and \ref{sec:field-dep_kernels}.
        } 
        To derive the corresponding FP equation, we apply It\^o's lemma to an arbitrary but fixed observable function $\mathcal{O} $, which yields the following stochastic differential equation
        \begin{align}
        \begin{split}
            d\mathcal{O} = &\left[\sum_{i = x, y} K^\gamma_i \, \partial_i \mathcal{O} 
            + \sum_{i, j = x, y} D_{ij} \, \partial_i \partial_j \mathcal{O} \right] d\theta
            \,+\, \\ &\,\, \sum_{i, j = x, y} \sqrt{2 N_i} \, (\partial_i \mathcal{O}) \, \sigma_{ij} \, dW_j(\theta),
        \end{split}
        \end{align}
        where the symmetric and positive-definite diffusion matrix is given by
        \begin{align}
            D = \frac{1}{2} \sigma^\top \sigma =
            \begin{pmatrix}
                N_x (\gamma_x^2 + \gamma_y^2) & \gamma_x \gamma_y \\
                \gamma_x \gamma_y & N_y (\gamma_x^2 + \gamma_y^2)
            \end{pmatrix}.
        \end{align}
        
        In direct analogy with the case without a kernel function, we obtain the FP equation associated with the kernelled CL process
        \begin{align} \label{eq:gamma_fpe}
            &\frac{d P^\gamma_{Z}}{d\theta}(x, y; \theta) \nonumber \\
            &=\sum_{i = x, y} \partial_i \left[ -K_i^\gamma \, P^\gamma_{Z}(x, y; \theta) + \partial_j \left( D_{ij} \, P^\gamma_{Z}(x, y; \theta) \right) \right] \nonumber \\
            &=: L^\top_{\gamma} \, P^\gamma_{Z}(x, y; \theta).
        \end{align}
        When acting on holomorphic observables, the operator $L_\gamma$ simplifies under the Cauchy–Riemann conditions. This motivates the corresponding complex FP equation for the complex weight $\rho(x, \theta) $, which takes the form
        \begin{align} \label{eq:cfpe_gamma}
            \frac{d\rho}{d\theta}(x, \theta) &= \partial_x \left[ - K^\gamma + \partial_x \gamma^2 \right] \rho(x, \theta) \nonumber \\
            &= \partial_x \left[ -\gamma^2 K + \gamma^2 \partial_x \right] \rho(x, \theta) \nonumber \\
            &=: L^\top_{c, \gamma} \, \rho(x, \theta),
        \end{align}
        and admits $\rho(x) = \exp[-S(x)]$ as a stationary solution. 
        
        \subsubsection{Criterion of correctness and properties for kernels}
        
        Based on the complex FP equation in Eq.~\eqref{eq:cfpe_gamma}, we can formulate an analogous correctness criterion for the kernelled case. When studying the time evolution in Eq.~\eqref{eq:tevolve} with the substitution $L^n \rightarrow L^n_\gamma = \gamma^2 (K +\partial_x)\partial_x $,%
        \footnote{We only consider kernel functions that ensure convergence of the integrals and permit partial integration, conditions that were implicitly assumed to hold in the case without a kernel. Care must be taken in the choice of kernel, as it can affect foundational assumptions of the CL method. In this work, we focus on polynomial kernels, whose well-behaved structure in the complex plane imposes useful constraints on both the design and optimization of the method.} 
        the dominant contributions again arise from integrals over powers of the modified drift magnitude $u_\gamma = \vert \gamma^2 K \vert $. Consequently, the associated drift magnitude density
        \begin{align}
            p_\gamma(u, \theta) = \int dx \, dy \, \delta\left(u_\gamma(x + i y) - u\right) P^\gamma_{Z}(x, y; \theta)
        \end{align}
        must exhibit at least exponential decay to ensure the convergence of the expansion. We note that in systems with multiple degrees of freedom or gauge symmetries, such as non-Abelian gauge theories, the definition of the norm used to evaluate the drift magnitude must be adapted accordingly. This issue is discussed in the context of SU($N$) lattice Yang–Mills theory in 3+1 dimensions in Sections~\ref{sec:coc_lattice} and \ref{sec:coc_kernels}.
        
        We emphasize that the introduction of a kernel function generally alters the real probability distribution $P $, and may thus enable the correctness criterion to be satisfied even in cases where it fails in the unkernelled setting. However, care must be taken when introducing kernel functions, as their zeros and singularities in the complex plane can pose serious problems. In particular, zeros of $\gamma$ can stall the dynamics, while singularities can destabilize the solution of the CL equation. To avoid such issues, we require that
        \begin{align}
            \begin{split}
                \int dx \, dy \, P^\gamma_{Z}(x, y; \theta) \, \delta\left( \left| \gamma(x + i y) \right| \right) &= \\
                \int dx \, dy \, P^\gamma_{Z}(x, y; \theta) \, \delta\left( \frac{1}{\left| \gamma(x + i y) \right|} \right) &= 0,
            \end{split}
        \end{align}
        for all Langevin times $\theta \geq \theta_0 $, as we have also discussed in Sec.~\ref{sec:coc_kernels}. In practice, we try to avoid such zeros and singularities in the design of kernels.

\newpage
\bibliographystyle{JHEP-2modlong}
\bibliography{ref}

\end{document}